\documentclass[letterpaper,pra,twocolumn,notitlepage,superscriptaddress, nofootinbib, showpacs, eqsecnum]{revtex4-1}

\usepackage[lmargin=2.2cm,rmargin=2.2cm,tmargin=3cm,bmargin=3cm] {geometry}
\usepackage {fancyhdr}
\usepackage{fancyvrb}
\usepackage [fleqn]{amsmath}
\usepackage {graphics}
\usepackage{setspace}
\usepackage{graphicx}
\usepackage{braket}
\usepackage{amsthm}
\usepackage{amssymb}
\usepackage{hanging}
\usepackage{url}
\usepackage{color}
\usepackage{mathtools}
\usepackage{tikz}
\usetikzlibrary{decorations}
\usetikzlibrary{patterns,snakes}

\providecommand{\abs}[1]{\lvert#1\rvert}

\newcommand{\blk}{\color{black}}

\DeclareMathOperator{\tr}{tr}
\DeclareMathOperator{\sech}{sech}
\DeclareMathOperator{\csch}{csch}


\usepackage{hyperref}
\usepackage{comment}

\begin{document}
\title{Noise analysis of single-qumode Gaussian operations using continuous-variable cluster states}
\author{Rafael N. Alexander} \email[Email:]{r.alexander@physics.usyd.edu.au}\affiliation{School of Physics, The University of Sydney, NSW, 2006, Australia}
\author{Seiji C. Armstrong}\affiliation{Department of Applied Physics, School of Engineering, The University of Tokyo, 7-3-1 Hongo, Bunkyo-ku, Tokyo 113-8656, Japan} \affiliation{Centre for Quantum Computation and Communication Technology, Department of Quantum Science, The Australian National University, Canberra, ACT 0200, Australia}
\author{Ryuji Ukai}\affiliation{Department of Applied Physics, School of Engineering, The University of Tokyo, 7-3-1 Hongo, Bunkyo-ku, Tokyo 113-8656, Japan}
 \author{Nicolas C. Menicucci} \affiliation{School of Physics, The University of Sydney, NSW, 2006, Australia}
\date{\today}
\begin{abstract}
We consider measurement-based quantum computation that uses scalable continuous-variable cluster states with a one-dimensional topology.  The physical resource, known here as the \emph{dual-rail quantum wire}, can be generated using temporally multiplexed offline squeezing and linear optics or by using a single optical parametric oscillator. We focus on an important class of quantum gates, specifically Gaussian unitaries that act on single modes, which gives universal quantum computation when supplemented with multi-mode operations and photon-counting measurements. The dual-rail wire supports two routes for applying single-qumode Gaussian unitaries: the first is to use traditional one-dimensional quantum-wire cluster-state measurement protocols. The second takes advantage of the dual-rail quantum wire in order to apply unitaries by measuring pairs of qumodes called \emph{macronodes}. We analyze and compare these methods in terms of the suitability for implementing single-qumode Gaussian measurement-based quantum computation. 
\end{abstract}
\maketitle

\section{Introduction}\label{sec:introduction}
The introduction of measurement-based quantum computation (MBQC) over a decade ago~\cite{aowqc} showed that adaptive local projective measurements alone are sufficient for quantum computation if a particular type of entangled resource called a \emph{cluster state}~\cite{peiaoip} is available. 
In the optical regime, the continuous-variable (CV) Gaussian analogue~\cite{cvgaocs, uqcwcvcs,Weedbrook:2012fe} of qubit cluster states can be generated deterministically~\cite{qcwcvc, bgcsblo} and in a highly scalable fashion~\cite{ugocvcs, owqcitofc, tofcaaowqc, alcvcsfasqng, tmcvcsulo,wqofcicvhcs, eromeo60moaqofc, ulscvcsmittd}. The generation of these states represents a big step towards achieving quantum computation using CVs~\cite{uqcwcvcs, qcwcvc}.

The canonical method for the construction of continuous-variable cluster states (CVCSs) uses momentum eigenstates~\cite{uqcwcvcs, qcwcvc}. In the optical setting, finitely squeezed states are typically used instead, as momentum eigenstates have infinite energy. Using squeezed states results in the construction of \emph{approximate CVCSs},\footnote{From here on we will assume that the term `CVCS' refers to the broader class of approximate physical states that approach the idealized case in the infinite squeezing limit.} which are intrinsically noisy. There is no way to eliminate this noise entirely~\cite{loqcwgcs, beicvcs, embqcwcvs}, but recent work has shown~\cite{ftmbqcwcvcs} that \emph{fault-tolerant} MBQC is possible using finitely squeezed CVCSs as long as qubit-based quantum information is appropriately encoded in the qumodes~\cite{Gottesman2001} and the level of squeezing in the cluster state (and encoded qubits) is above a fixed, finite value called the~\emph{squeezing threshold}.

Related to this issue is the development of methods for MBQC that attempt to use available experimental squeezing resources more efficiently, in the sense that they introduce less noise from finite squeezing. Improving these methods will help to reduce the experimental demands set by an error-corrective approach for dealing with finite squeezing.  One approach is to optimize the CVCS generation process to produce better-quality approximations of ideal CVCSs from the available resources~\cite{bgcsblo}.

An additional concern for top-down approaches is the trade-off between the quality of the approximation and the scalability of the construction process. Methods for generating CVCSs that employ optical parametric oscillators (OPOs) have shown excellent scalability. Two examples are the \emph{single-OPO method}~\cite{ugocvcs,  owqcitofc, tofcaaowqc} and the \emph{temporal-mode linear optics method}~\cite{tmcvcsulo}.

 The single-OPO method generates entangled states in a single-shot, using an OPO. All of the squeezing and linear optics takes place inside the OPO, and the OPO cavity eigenmodes serve as the carriers of quantum information, referred to here as \emph{qumodes}~\cite{ugocvcs, owqcitofc, tofcaaowqc, pgoqceitofc, epoqcaghzesfcv,  pgoqceitofc}. This method sets demands on the OPO (specifically the nonlinear crystal contained within it) and the frequency content of the pump beam. Once these prerequisites can be achieved within the laboratory~\cite{eromeo60moaqofc}, CVCSs can be generated with a pump beam complexity that scales as a \emph{constant} with the number of qumodes. By using multiple OPOs, this method can be used to generate higher-dimensional graph structures~\cite{wqofcicvhcs}. 

The temporal-mode linear optics method works by generating a small section of the cluster state and then repeatedly extending it (as required) using a basic set of optical machinery~\cite{alcvcsfasqng}. It uses temporally encoded qumodes, offline squeezing, and linear optics to generate the cluster~\cite{tmcvcsulo}. In a recent result, this was achieved on the scale of over $10,000$ entangled qumodes~\cite{ulscvcsmittd}.
Both of these methods can generate states with 1D and 2D topologies~\cite{tmcvcsulo}. We call these the \emph{dual-rail quantum wire}~(DRW)\footnote{Not to be confused with dual-rail photonic qubits, as in Ref.~\cite{loqcwpq}. } and the \emph{quad-rail lattice}, respectively. 

This work provides a basic framework for characterizing quantum computation on the DRW, showing how the noise introduced to the computation by finite squeezing depends on the measurement protocol used to implement gates. We will focus on the set of unitaries that can be implemented using just homodyne detection on the DRW. This set is an important subgroup of all single-qumode unitaries: single-qumode \emph{Gaussian} unitaries~\cite{qcwcvc, Weedbrook:2012fe}. 
Adding the ability to count photons enables universal single-qumode MBQC on the DRW. 

Extending this to universal quantum computation requires supplementing the above resources with a multi-qumode gate. Some results in this direction involve introducing additional linear cluster-state resources and Bell measurements in order to apply entangling gates between pairs of qumodes \cite{msmiowqipwsatmol}.

Alternatively, one can use a CVCS with higher-dimensional graph structure to perform a two-qumode gate using measurements alone. On a CVCS with 2D topology, such as the quad-rail lattice, homodyne detection alone implements all multi-qumode Gaussian unitaries, and the addition of photon counting enables fully universal MBQC~\cite{qcwcvc}. Our analysis will be limited to the DRW, with generalization to the quad-rail lattice left to future work. 

\begin{figure}
%
\begin{tikzpicture}[scale=1]
\path (0,0) coordinate (origin);
\path (1,1) coordinate (P0);
\path (1,0) coordinate (P1);
\path (2,1) coordinate (P2);
\path (2,0) coordinate (P3);
\path (3,1) coordinate (P4);
\path (3,0) coordinate (P5);
\path (4,1) coordinate (P6);
\path (0.5,0.5) coordinate (t1);
\path (1.5,0.5) coordinate (t2);
\path (2.5,0.5) coordinate (t3);
\path (3.5,0.5) coordinate (t4);
\path (0,1) coordinate (E1);
\path (-0.5, 0.5) coordinate (E2);
\path (4, 0) coordinate (E3);
\path (4.5, 0.5) coordinate (E4);
\node at (-1,1.5) {(a)};
\path (P2) edge[ out=120, in=60, looseness=1, loop, distance=0.35cm, color = red, line width=0.2mm]
            node[above=3pt] {
} (P2);
\path (P3) edge[ out=-120, in=-60, looseness=1, loop, distance=0.35cm, color = red, line width=0.2mm]
            node[below=3pt] {
} (P3);
\path (P4) edge[ out=120, in=60, looseness=1, loop, distance=0.35cm, color =red, line width=0.2mm]
            node[above=3pt] {
} (P4);
\path (P5) edge[ out=-120, in=-60, looseness=1, loop, distance=0.35cm, color = red, line width=0.2mm]
            node[below=3pt] {
} (P5);
\path (P0) edge[ out=120, in=60, looseness=1, loop, distance=0.35cm, color =red, line width=0.2mm]
            node[above=3pt] {
} (P0);
\path (P1) edge[ out=-120, in=-60, looseness=1, loop, distance=0.35cm, color =red, line width=0.2mm]
            node[below=3pt] {
} (P1);
\path (P6) edge[ out=120, in=60, looseness=1, loop, distance=0.35cm, color =red, line width=0.2mm]
            node[above=3pt] {
} (P6);
\path (origin) edge[ out=-120, in=-60, looseness=1, loop, distance=0.35cm, color = red, line width=0.2mm]
            node[below=3pt] {
} (origin);
\path (E1) edge[ out=120, in=60, looseness=1, loop, distance=0.35cm, color = red, line width=0.2mm]
            node[below=3pt] {} (E1);
\path (E3) edge[ out=-120, in=-60, looseness=1, loop, distance=0.35cm, color = red, line width=0.2mm]
            node[below=3pt] {} (E3);  
 \begin{scope}[color=blue,line width=0.2mm]
\draw (origin) -- (P0) (P1) -- (P2) (P3) -- (P4) (P5) -- (P6);
\end{scope}
 \begin{scope}[color=blue,line width=0.2mm]
\draw (origin) -- (P1) (P1) -- (P3) (P3) -- (P5) (P5) -- (E3);
\end{scope}
 \begin{scope}[color=orange,line width=0.2mm]
\draw (P0) -- (P2) (P2) -- (P4) (P4) -- (P6) (P0)--(P3) (P2)--(P5) (E1)--(P0) (E1)--(P1) (P4)--(E3);
\end{scope}
 \begin{scope}[color=blue,line width=0.2mm, dashed]
\draw (E1) -- (E2) (E3) -- (E4)  (-0.5, 0) -- (origin) (E3)-- (4.5, 0);
\end{scope}
 \begin{scope}[color=orange,line width=0.2mm, dashed]
\draw  (-0.5, 1)--(E1) (E2)--(origin) (P6)--(E4) (P6)--(4.5, 1);
\end{scope}
 \shade[ball color=black] (origin) circle(0.05);
 \shade[ball color=black] (P0) circle(0.05);
 \shade[ball color=black] (P1) circle(0.05);
 \shade[ball color=black] (P2) circle(0.05);
 \shade[ball color=black] (P3) circle(0.05);
 \shade[ball color=black] (P4) circle(0.05);
\shade[ball color=black] (P5) circle(0.05);
\shade[ball color=black] (P6) circle(0.05);
\shade[ball color=black] (E1) circle(0.05);
\shade[ball color=black] (E3) circle(0.05);
%
\end{tikzpicture}
%
%
\begin{tikzpicture}[scale=1]
\fill[white] (0,0) rectangle (4,0.5);
\end{tikzpicture}
\vspace{-1.5cm}
%
\begin{tikzpicture}[scale=1]
\hspace{-4.6cm}
\fill [color=green, opacity=0.25]  (7 ,0) ellipse (0.15 and 0.25);
\fill [color=blue, opacity=0.25]  (8 ,0) ellipse (0.15 and 0.25);
\fill [color=blue, opacity=0.25]  (9 ,0) ellipse (0.15 and 0.25);
\fill [color=blue, opacity=0.25]  (10 ,0) ellipse (0.15 and 0.25);
\fill [color=blue, opacity=0.25]  (11, 0) ellipse (0.15 and 0.25);
\fill [color=blue, opacity=0.25]  (7, 1) ellipse (0.15 and 0.25);
\fill [color=blue, opacity=0.25]  (8 ,1) ellipse (0.15 and 0.25);
\fill [color=blue, opacity=0.25]  (9 ,1) ellipse (0.15 and 0.25);
\fill [color=blue, opacity=0.25]  (10 ,1) ellipse (0.15 and 0.25);
\fill [color=blue, opacity=0.25]  (11 ,1) ellipse (0.15 and 0.25);
\draw[line width=0.5mm, ->, color=purple]  node[label={[xshift=6.6cm, yshift=-0.1cm]$\color{black}\hat{b}_{1}\color{blue}$}] {} (6.75,1.25) -- (6.9,1.1);
\draw[line width=0.5mm, ->, color=purple] node[label={[xshift=7.6cm, yshift=-0.1cm]$\color{black}\hat{b}_{2}\color{blue}$}] {} (7.75,1.25) -- (7.9,1.1);
\draw[line width=0.5mm, ->, color=purple] node[label={[xshift=8.6cm, yshift=-0.1cm]$\color{black}\hat{b}_{3}\color{blue}$}] {} (8.75,1.25) -- (8.9,1.1);
\draw[line width=0.5mm, ->, color=purple] node[label={[xshift=9.6cm, yshift=-0.1cm]$\color{black}\hat{b}_{4}\color{blue}$}] {} (9.75,1.25) -- (9.9,1.1);
\draw[line width=0.5mm, ->, color=purple] node[label={[xshift=10.6cm, yshift=-0.1cm]$\color{black}\hat{b}_{5}\color{blue}$}] {} (10.75,1.25) -- (10.9,1.1);
\draw[line width=0.5mm, ->, color=purple] node[label={[xshift=6.6cm, yshift=0.9cm]$\color{black}\hat{q}\color{blue}$}] {} (6.75,0.25) -- (6.9,0.1);
\draw[line width=0.5mm, ->, color=purple] node[label={[xshift=7.6cm, yshift=0.9cm]$\color{black}\hat{q}\color{blue}$}] {} (7.75,0.25) -- (7.9,0.1);
\draw[line width=0.5mm, ->, color=purple] node[label={[xshift=8.6cm, yshift=0.9cm]$\color{black}\hat{q}\color{blue}$}] {} (8.75,0.25) -- (8.9,0.1);
\draw[line width=0.5mm, ->, color=purple] node[label={[xshift=9.6cm, yshift=0.9cm]$\color{black}\hat{q}\color{blue}$}] {} (9.75,0.25) -- (9.9,0.1);
\draw[line width=0.5mm, ->, color=purple] node[label={[xshift=10.6cm, yshift=0.9cm]$\color{black}\hat{q}\color{blue}$}] {} (10.75,0.25) -- (10.9,0.1);
\node at (6,1.5) {(b)};
\path (7,0) coordinate (origin);
\path (8,0) coordinate (P1);
\path (9,0) coordinate (P3);
\path (10,0) coordinate (P5);
\path (11, 0) coordinate (E3);
\path (P3) edge[ out=-120, in=-60, looseness=1, loop, distance=0.35cm, color = red, line width=0.2mm]
            node[below=3pt] {
} (P3);
\path (P5) edge[ out=-120, in=-60, looseness=1, loop, distance=0.35cm, color = red, line width=0.2mm]
            node[below=3pt] {
} (P5);
\path (P1) edge[ out=-120, in=-60, looseness=1, loop, distance=0.35cm, color = red, line width=0.2mm]
            node[below=3pt] {
} (P1);
\path (origin) edge[ out=-120, in=-60, looseness=1, loop, distance=0.35cm, color = black, line width=0.2mm]
            node[below=3pt] {
} (origin);
\path (E3) edge[ out=-120, in=-60, looseness=1, loop, distance=0.35cm, color =red, line width=0.2mm]
            node[below=3pt] {
} (E3);
 \begin{scope}[color=blue,line width=0.2mm]
\draw (origin) -- (P1) (P1) -- (P3) (P3) -- (P5) (P5) -- (E3);
\end{scope}
 \begin{scope}[color=blue,line width=0.2mm, dashed]
\draw  (E3)-- (11.5, 0);
\end{scope}
 \shade[ball color=black] (7,0) circle(0.05);
 \shade[ball color=black] (7,1) circle(0.05);
 \shade[ball color=black] (8, 0) circle(0.05);
 \shade[ball color=black] (8,1) circle(0.05);
 \shade[ball color=black] (9,0) circle(0.05);
 \shade[ball color=black] (9,1) circle(0.05);
\shade[ball color=black] (10,0) circle(0.05);
\shade[ball color=black] (10,1) circle(0.05);
\shade[ball color=black] (11,0) circle(0.05);
\shade[ball color=black] (11,1) circle(0.05);
%
\end{tikzpicture}
%
\begin{tikzpicture}[scale=1]
\hspace{0.00cm}
\fill [color=green, opacity=0.25]  (0 ,-3) ellipse (0.30 and 0.8);
\fill [color=blue, opacity=0.25]  (1,-3) ellipse (0.30 and 0.8);
\fill [color=blue, opacity=0.25]  (2 ,-3) ellipse (0.30 and 0.8);
\fill [color=blue, opacity=0.25]  (3,-3) ellipse (0.30 and 0.8);
\fill [color=blue, opacity=0.25]  (4,-3) ellipse (0.30 and 0.8);
\path (0, -3.5) coordinate (origin);
\path (1,-2.5) coordinate (P0);
\path (1, -3.5) coordinate (P1);
\path (2,-2.5) coordinate (P2);
\path (2, -3.5) coordinate (P3);
\path (3,-2.5) coordinate (P4);
\path (3,-3.5) coordinate (P5);
\path (4,-2.5) coordinate (P6);
\path (0.5, -3) coordinate (t1);
\path (1.5, -3) coordinate (t2);
\path (2.5, -3) coordinate (t3);
\path (3.5, -3) coordinate (t4);
\path (0, -2.5) coordinate (E1);
\path (0.5, -3) coordinate (E2);
\path (4, -3.5) coordinate (E3);
\path (4.5, -3) coordinate (E4);
\node at (-1,-1.75) {(c)};
\path (P2) edge[ out=120, in=60, looseness=1, loop, distance=0.35cm, color =red, line width=0.2mm]
            node[above=3pt] {
} (P2);
\path (P3) edge[ out=-120, in=-60, looseness=1, loop, distance=0.35cm, color =red, line width=0.2mm]
            node[below=3pt] {
} (P3);
\path (P4) edge[ out=120, in=60, looseness=1, loop, distance=0.35cm, color =red, line width=0.2mm]
            node[above=3pt] {
} (P4);
\path (P5) edge[ out=-120, in=-60, looseness=1, loop, distance=0.35cm, color = red, line width=0.2mm]
            node[below=3pt] {
} (P5);
\path (P0) edge[ out=120, in=60, looseness=1, loop, distance=0.35cm, color = red, line width=0.2mm]
            node[above=3pt] {
} (P0);
\path (P1) edge[ out=-120, in=-60, looseness=1, loop, distance=0.35cm, color =red, line width=0.2mm]
            node[below=3pt] {
} (P1);
\path (P6) edge[ out=120, in=60, looseness=1, loop, distance=0.35cm, color = red, line width=0.2mm]
            node[above=3pt] {
} (P6);
\path (origin) edge[ out=-120, in=-60, looseness=1, loop, distance=0.35cm, color = black, line width=0.2mm]
            node[below=3pt] {
} (origin);
\path (E1) edge[ out=120, in=60, looseness=1, loop, distance=0.35cm, color =black, line width=0.2mm]
            node[above=3pt] {
} (E1);
\path (E3) edge[ out=-120, in=-60, looseness=1, loop, distance=0.35cm, color =red, line width=0.2mm]
            node[below=3pt] {
} (E3);
 \begin{scope}[color=blue,line width=0.2mm]
\draw (origin) -- (P0) (P1) -- (P2) (P3) -- (P4) (P5) -- (P6) (origin)--(P1) (P1)--(P3) (P3)--(P5) (P5)--(E3);
\end{scope}
 \begin{scope}[color=blue,line width=0.2mm, dashed]
\draw (E3) -- (E4) (E3)--(4.5, -3.5);
\end{scope}
 \begin{scope}[color=orange,line width=0.2mm, dashed]
\draw (4, -2.5) -- (E4) (4, -2.5)--(4.5, -2.5);
\end{scope}
 \begin{scope}[color=orange,line width=0.2mm]
\draw (P0) -- (P2) (P2) -- (P4) (P4) -- (P6) (P0)--(P3) (P2)--(P5) (E1)--(P0) (E1)--(P1) (P4)--(E3);
\end{scope}
 \begin{scope}[color=black,line width=0.2mm]
\draw (0, -2.5) -- (0, -3.5);
\end{scope}
 \shade[ball color=black] (origin) circle(0.05);
 \shade[ball color=black] (P0) circle(0.05);
 \shade[ball color=black] (P1) circle(0.05);
 \shade[ball color=black] (P2) circle(0.05);
 \shade[ball color=black] (P3) circle(0.05);
 \shade[ball color=black] (P4) circle(0.05);
\shade[ball color=black] (P5) circle(0.05);
\shade[ball color=black] (P6) circle(0.05);
\shade[ball color=black] (E1) circle(0.05);
\shade[ball color=black] (E3) circle(0.05);
\draw[line width=0.5mm, ->, color=purple]  node[label={[xshift=-0.25cm, yshift=-4.35cm]$\color{black}\hat{b}_{1b}\color{blue}$}] {} (-0.25,-3.85) -- (-0.10,-3.7);
\draw[line width=0.5mm, ->, color=purple]  node[label={[xshift=0.75cm, yshift=-4.35cm]$\color{black}\hat{b}_{2b}\color{blue}$}] {} (0.75,-3.85) -- (0.9,-3.7);
\draw[line width=0.5mm, ->, color=purple]  node[label={[xshift=1.75cm, yshift=-4.35cm]$\color{black}\hat{b}_{3b}\color{blue}$}] {} (1.75,-3.85) -- (1.9,-3.7);
\draw[line width=0.5mm, ->, color=purple]  node[label={[xshift=2.75cm, yshift=-4.35cm]$\color{black}\hat{b}_{4b}\color{blue}$}] {} (2.75,-3.85) -- (2.9,-3.7);
\draw[line width=0.5mm, ->, color=purple]  node[label={[xshift=3.75cm, yshift=-4.35cm]$\color{black}\hat{b}_{5b}\color{blue}$}] {} (3.75,-3.85) -- (3.9,-3.7);
\draw[line width=0.5mm, ->, color=purple]  node[label={[xshift=-0.25cm, yshift=-2.4cm]$\color{black}\hat{b}_{1a}\color{blue}$}] {} (-0.25,-2.15) -- (-0.10,-2.3);
\draw[line width=0.5mm, ->, color=purple]  node[label={[xshift=0.75cm, yshift=-2.4cm]$\color{black}\hat{b}_{2a}\color{blue}$}] {} (0.75,-2.15) -- (0.9,-2.3);
\draw[line width=0.5mm, ->, color=purple]  node[label={[xshift=1.75cm, yshift=-2.4cm]$\color{black}\hat{b}_{3a}\color{blue}$}] {} (1.75,-2.15) -- (1.9,-2.3);
\draw[line width=0.5mm, ->, color=purple]  node[label={[xshift=2.75cm, yshift=-2.4cm]$\color{black}\hat{b}_{4a}\color{blue}$}] {} (2.75,-2.15) -- (2.9,-2.3);
\draw[line width=0.5mm, ->, color=purple]  node[label={[xshift=3.75cm, yshift=-2.4cm]$\color{black}\hat{b}_{5a}\color{blue}$}] {} (3.75,-2.15) -- (3.9,-2.3);
\end{tikzpicture}
\caption{(Color online) Two ways of implementing measurement-based quantum computation on the dual-rail quantum wire~(DRW). $\mathbf{(a)}$ Simplified graph~\cite{tmcvcsulo} of the DRW. $\mathbf{(b)}$ The continuous-variable quantum-wire~(CVW) protocol involves converting the DRW to a CVW by measurement of the position-quadrature ($\hat{q}$) basis on the top qumodes~\cite{tmcvcsulo}, followed by single-qumode homodyne measurements in some quadrature bases $\hat{b}_{i}$ to evolve and propagate the state to the right along the wire~\cite{qcwcvc, eogcc, ulbttowqc}. $\mathbf{(c)}$ The macronode protocol involves encoding the input state within the leftmost macronode (pair of qumodes). Each macronode is measured by homodyne detection of its constituent qumodes ($\hat{b}_{ia}$, $\hat{b}_{ib}$). Graph weights~\cite{gcfgps} have been omitted for convenience. Color represents the sign/phase of each link. Blue/orange represents $\pm$ sign respectively, and red is a complex sign of~$i$. The magnitude of each red self-loop is $\varepsilon_{\text{D}} = \sech{(2\alpha)}$ where $\alpha >0$ is the overall squeezing parameter~\cite{tmcvcsulo}. The adjoining edges have magnitude $g_{\text{D}}=\frac{1}{2}\tanh{(2\alpha)}$. The black edge and self-loops contained in the green ovals label the modes containing the encoded input state.} \label{fig:dualsummary}
\end{figure}
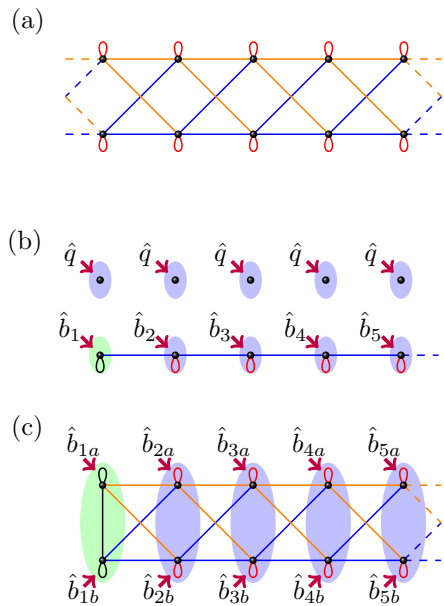

In particular, we consider two measurement protocols. The first applies traditional continuous-variable quantum wire (CVW) cluster measurements~\cite{qcwcvc, eogcc, ulbttowqc} to the state, using the fact that the DRW can be converted to a CVW. We will refer to this protocol as the \emph{CVW protocol}. The other type treats the DRW as a double-thick quantum wire with pairs of nodes called \emph{macronodes} at each wire site. We refer to this protocol as the \emph{macronode protocol}. It bears some resemblance to sequential CV teleportation~\cite{ulscvcsmittd}. These approaches are illustrated in Fig.~\ref{fig:dualsummary}.

Our analysis of the CVW protocol involves consideration of a class of CVWs containing states generated by methods of interest discussed above. This class is characterized by just two parameters, which are weights that label the edges of the CVW graph~\cite{gcfgps}. We relate these graphical parameters to noise introduced by finite squeezing during single-qumode Gaussian quantum computation, showing that, despite scalability inherited from the DRW generation process, the CVW protocol is a suboptimal strategy because the values of the graphical variables for the CVW introduce excessive noise to MBQC.

The key feature of the macronode protocol is that it does not involve conversion of the DRW into a CVW and thereby makes full use of the available squeezing. We show that this type of protocol can be used to implement arbitrary single-qumode Gaussian unitary gates using fewer qumodes than the CVW protocol.  We also discuss an interesting special case of the macronode protocol that allows us to reproduce a CVW-like mode of computation, which we call here the \emph{dictionary protocol}. This protocol allows us to port measurement procedures (and hence, algorithms) that apply to the CVW directly to the DRW. We show that the noise properties of the general macronode protocol are more favorable than both the dictionary and the CVW protocols, which both perform comparably in terms of noise.


The structure of this paper is as follows: In Sec.~\hyperref[sec:CVquantumwire]{\ref{sec:CVquantumwire}} we review single-qumode Gaussian quantum computation on a class of CVWs known as \emph{uniformly weighted wires}. Next we quantify the noise introduced in a computation due to finite squeezing for the CVW protocol. In Sec.~\hyperref[sec:dualrail]{\ref{sec:dualrail}} we introduce the macronode and dictionary protocols. We compare them to each other and to the CVW protocol, with respect to noise per single-qumode Gaussian unitary, showing that the macronode protocol always outperforms the others. As a quantitative application of our results, in  Sec.~\hyperref[sec:numbermeasurements]{\ref{sec:numbermeasurements}} we analyze the noise of implementing rotation gates using three or four measurements in the different protocols. We show that the extra degree of freedom in the four-measurement case can lead to a dramatic reduction in the noise for particular gates, while three measurements remain favorable for others.
%
%
%
%
%
%
%
%
%
%
%
%
Section~\hyperref[sec:discussionandconclusion]{\ref{sec:discussionandconclusion}} concludes with some discussion. \blk

\section{CVW protocol}\label{sec:CVquantumwire}
Traditionally, the connection between CVCSs and graphs is as follows~\cite{qcwcvc}. Nodes/vertices represent momentum eigenstates, and weighted links/edges between them represent the application of a \text{controlled-Z} gate (defined below) between two qumodes, with the gate interaction strength being equal to the edge weight (usually the weight of each graph edge is 1). As a graphical description of CVCSs, this is unphysical because the corresponding states cannot be normalized. However, it can be taken as the infinite-squeezing limit of approximate CVCSs, which are Gaussian pure states~\cite{gcfgps}.  Given a particular ideal CVCS with adjacency matrix $\mathbf{A}$, the corresponding family of approximate CVCSs is defined by those states whose complex graph $\mathbf{Z}$~\cite{gcfgps} approaches $\mathbf{A}$ in the infinite-squeezing limit. Since $\mathbf{A}$ has real entries, the imaginary part of $\mathbf{Z}$ must vanish in this limit. 

We will consider a restricted class of approximate CVCSs that are CVWs with uniformly weighted graphs of the following form~\cite{gcfgps}:
\begin{equation}
\mathbf{Z}= g\mathbf{A}_{\text{BL}} + i \varepsilon \mathbf{I}. \label{eq:graphclasslinearcvcs}
\end{equation}
Here $\mathbf{A}_{\text{BL}}$ is a binary (B) adjacency matrix corresponding to a linear (L) graph. The parameter $g$ is allowed to take any real value (it comes from the controlled-Z gate; see Eq.~\eqref{eq:czgate}) and is assigned to $all$ the links between neighbouring nodes on the graph. The second term describes the self-loop edges, which all have weight $\varepsilon$, and we require that $\varepsilon\rightarrow 0$ in the infinite-squeezing limit. For CVWs produced from the DRW, these weights are denoted as $g_{\text{D}}$ and $\varepsilon_{\text{D}}$ and have a specific form that depends on the overall squeezing parameter, $\alpha >0$~\cite{tmcvcsulo} (see Fig.~\ref{fig:dualsummary}). They are defined as
\begin{equation}
g_{\text{D}} \coloneqq \frac{1}{2}\tanh{(2\alpha )} \label{eq:gdual}
\end{equation}
 and 
\begin{equation}
\varepsilon_{\text{D}} \coloneqq \sech{(2\alpha )}. \label{eq:edual}
\end{equation}
We will assume we are working in the general uniformly-weighted wire case, except for when drawing conclusions specifically for CVWs produced from the DRW. We use the notation $\{ g$, $\varepsilon \}$ and $\{ g_{\text{D}}$, $\varepsilon_{\text{D}}\}$, respectively, in order to distinguish these cases.

\subsection{MBQC on uniformly-weighted CVWs}
Once a suitable CVW resource state has been generated, single-qumode Gaussian computation proceeds by measuring linear combinations of the canonical position and momentum quadrature operators---$\hat{q}$ and $\hat{p}$, respectively---on nodes on the wire. We employ the conventions that $[\hat q, \hat p] = i$ and $\hbar = 1$, which means that the variance of a qumode in its vacuum state is always $\langle \hat q^2 \rangle_{\text{vac}} = \langle \hat p^2 \rangle_{\text{vac}} = \tfrac 1 2$. The particular measurements in question will be
\begin{align}
\label{eq:linearbasis}
\hat{b}_{i} \coloneqq  \alpha_{i} \bigl[ (\cos{\theta_{i}}) \hat{p} + (\sin{\theta_{i}}) \hat{q} \bigr]  = \hat{p}+ \sigma_{i} \hat{q},
\end{align}
where $\sigma_{i} = \tan{\theta_{i}}$ and $\alpha_{i}= \sec{\theta_{i}}$. Performing the logical measurement $\hat{b}_{i}$ is equivalent to physically measuring the rotated quadrature operator $(\cos{\theta_{i}}) \hat{p} + (\sin{\theta_{i}}) \hat{q}$ and then multiplying the measurement outcome by $\alpha_{i}$; such measurements can be achieved experimentally through homodyne detection~\cite{uqcwcvcs, ulbttowqc}. Below, we describe the effect of these measurements on an input state, but first we define some standard single- and two-qumode Gaussian operations~\cite{qcwcvc}.

%
%
The controlled-$Z$ gate is
\begin{equation} \label{eq:czgate}
\hat{C}_{Z} (g) \coloneqq \exp\left(i g \hat{q} \otimes \hat{q} \right),
\end{equation}
where $g$ is the interaction strength \cite{gcfgps}. It is a two-qumode entangling gate.
The single-qumode squeezing gate, which squeezes the $\hat q$ quadrature by a factor of~$s$ (called the \emph{squeezing factor}), is
\begin{equation}
\hat{S}(s) \coloneqq \exp\left[-i \left(\frac {\ln s}{2}\right) (\hat{q}\hat{p}+\hat{p}\hat{q}) \right],\label{eq:squeezegate}
\end{equation}
where $\ln s$ is called the \emph{squeezing parameter}. We represent its Heisenberg action on the vector of single-qumode quadrature operators $\mathbf{\hat{x}}=(\hat{q}, \hat{p})^{T}$ by the symplectic matrix $\mathbf{S}(s)$:
\begin{equation}
\hat{S}^{\dagger} (s) \hat{\mathbf{x}} \hat{S}(s) = \mathbf{S}(s) \mathbf{\hat{x}}=\begin{pmatrix} s & 0 \\ 0 & s^{-1} \end{pmatrix} \begin{pmatrix} \hat{q} \\ \hat{p} \end{pmatrix}.
\end{equation}
In the Heisenberg picture, this operator has the action of rescaling the position and momentum quadratures by $s$ and $s^{-1}$ respectively.
 
The shearing gate is defined as
\begin{equation}
\hat{P} (\sigma ) \coloneqq \exp\left(\frac {i \sigma \hat{q}^{2}}{2} \right),
\end{equation}
with $\sigma$ called the \emph{shearing parameter}. We represent its Heisenberg action on $\mathbf{\hat{x}}$ by the symplectic matrix $\mathbf{P}(\sigma )$:
\begin{align}
\hat{P}^{\dagger} (\sigma ) \hat{\mathbf{x}} \hat{P}(\sigma ) = \mathbf{P}(\sigma ) \mathbf{\hat{x}}=\begin{pmatrix} 1 & 0 \\ \sigma & 1 \end{pmatrix} \begin{pmatrix} \hat{q} \\ \hat{p} \end{pmatrix}.
\end{align}
In the Heisenberg picture, this operator acts as a shear in phase space parallel to the momentum axis by a gradient $\sigma$.

The Fourier transform is
\begin{equation}
\hat{F} \coloneqq \exp\left[\frac{ i\pi}{4} \left( \hat{q}^{2}+\hat{p}^{2}\right) \right]
\end{equation}
We represent its Heisenberg action on $\mathbf{\hat{x}}$ by the symplectic matrix $\mathbf{F}$:
\begin{equation}
\hat{F}^{\dagger}\hat{\mathbf{x}} \hat{F}= \mathbf{F} \mathbf{\hat{x}}=\begin{pmatrix} 0 & -1 \\ 1 & 0 \end{pmatrix} \begin{pmatrix} \hat{q} \\ \hat{p} \end{pmatrix}.
\end{equation}
In the Heisenberg picture, this operator acts as a $\frac{\pi}{2}$ clockwise rotation of the quadratures. This is a special case of a more general rotation,
\begin{equation}
\hat{R}(\theta) \coloneqq \exp{\left[ \frac{i \theta}{2} (\hat{q}^{2}+ \hat{p}^{2})\right]}, \label{eq:rotationgate}
\end{equation}
whose Heisenberg action on $\mathbf{\hat{x}}$ is given by the symplectic matrix $\mathbf{R}(\theta )$:
\begin{equation}
\hat{R}^{\dagger}(\theta )\hat{\mathbf{x}} \hat{R} (\theta )= \mathbf{R}(\theta ) \mathbf{\hat{x}}=\begin{pmatrix} \cos\theta & -\sin\theta \\ \sin\theta & \cos\theta \end{pmatrix} \begin{pmatrix} \hat{q} \\ \hat{p} \end{pmatrix}.
\end{equation}
In the Heisenberg picture, this operator rotates the quadrature operators clockwise by an angle~$\theta$. These gates will be useful throughout the rest of this Article.

\begin{figure}
\begin{center}
    %
\begin{tikzpicture}[thick]
    %
    \tikzstyle{operator} = [draw,fill=white,minimum size=1.5em] 
    \tikzstyle{phase} = [fill,shape=circle,minimum size=5pt,inner sep=0pt]
    %
    \node at (-0.1,0) (q1) {$\ket{\psi}$};
    \node at (-0.1,-1) (q2) {$\ket{0}$};
    %
    \node[operator] (op21) at (1.2,-1) {S$\left(\varepsilon^{-1/2}\right)$} edge [-] (q2);
    %
    \node[phase] (phase11) at (2.3,0) {} edge [-] (q1);
    \node[phase] (phase12) at (2.3,-1) {} edge [-] (op21);
    \draw[-] (phase11) -- (phase12);
\node at (2.5, -0.5) {$g$};
    %
    %
\draw (3.6, 0.25)-- (3.1,0.25)--(3.1, -0.25)--(3.6,-0.25);
\draw (3.6,-0.25) arc(-90:90:0.25) [thick];
    \node at (3.4,0) (pm) {$\hat{b}_{1}$}; 
   \node (end1) at (3.2,0) {} edge [-] (q1);
    \node (end2) at (3.2,-1) {} edge [-] (phase12);
\node at (3.75,0) {} edge [-,double] (4.2,0);
\node at (4.45,-0.05) {$m_{1}$};
\node at (4.7,-1) {$\hat{N}(\varepsilon ) \hat{X} \left( \frac{m_{1}}{g} \right)  \hat{U}_{1} \ket{\psi}$};
\filldraw (2.6, -1) circle (2pt);
\draw[-] (2.6, -1)--(2.6, -1.4);
\draw[dotted] (2.6, -1.4)--(2.6, -1.7);
\draw[dotted] (-0.5, -1.4)--(-0.5, -1.7);
\draw[dotted] (2.8, -1.4)--(2.8, -1.7);
\draw[dotted] (-0.5, -1.4) -- (-0.5, -0.5)--(1.7, -0.5)--(1.7, 0.3)--(2.8, 0.3)--(2.8, -1.4);
\path (0,-3) coordinate (origin);
\path (1,-3) coordinate (P1);
\path (2,-3) coordinate (P3);
\path (3,-3) coordinate (P5);
\path (4, -3) coordinate (E3);
\path (0.5, -3) node[label={[shift={(0,-0.65)}]$g$}] {};
\path (1.5, -3) node[label={[shift={(0,-0.65)}]$g$}] {};
\path (2.5, -3) node[label={[shift={(0,-0.65)}]$g$}] {};
\path (3.5, -3) node[label={[shift={(0,-0.65)}]$g$}] {};
\path (P3) edge[ out=120, in=60, looseness=1, loop, distance=0.35cm, color = purple, line width=0.2mm]
            node[above=3pt] {
$\color{black}i\varepsilon\color{blue}$ 
} (P3);
\path (P5) edge[ out=120, in=60, looseness=1, loop, distance=0.35cm, color = purple, line width=0.2mm]
            node[above=3pt] {
$\color{black}i\varepsilon\color{blue}$
} (P5);
\path (P1) edge[ out=120, in=60, looseness=1, loop, distance=0.35cm, color = blue, line width=0.2mm]
            node[above=3pt] {
$\color{black}i\varepsilon\color{blue}$
} (P1);
\path (origin) edge[ out=120, in=60, looseness=1, loop, distance=0.35cm, color = blue, line width=0.2mm]
            node[above=3pt] {
$\color{black}z_{\psi}\color{blue}$
} (origin);
\path (E3) edge[ out=120, in=60, looseness=1, loop, distance=0.35cm, color = purple, line width=0.2mm]
            node[above=3pt] {$\color{black}i\varepsilon\color{blue}$} (E3);
 \begin{scope}[color=blue,line width=0.2mm]
\draw (origin) -- (P1) (P1) -- (1.5,-3);
\end{scope}
 \begin{scope}[color=purple,line width=0.2mm, dashed]
\draw (1.5, -3) -- (P3) (P3) -- (P5) (P5) -- (E3);
\end{scope}
 \begin{scope}[color=purple,line width=0.2mm, dashed]
\draw  (E3)-- (4.5, -3);
\end{scope}
 \shade[ball color=black] (0,-3) circle(0.05);
 \shade[ball color=black] (1, -3) circle(0.05);
  \shade[ball color=black] (2,-3) circle(0.05);
\shade[ball color=black] (3,-3) circle(0.05);
\shade[ball color=black] (4,-3) circle(0.05);
\node at (-1,0.5) {(a)};
\node at (-1,-2.5) {(b)};
      \end{tikzpicture}
\end{center}
\caption{(Color online) $\mathbf{(a)}$ Circuit diagram for an element of Gaussian measurement-based quantum computation. The input $\ket{\psi}$ is entangled with a momentum-squeezed vacuum state by a $\hat{C}_{Z} (g)$ interaction (see Eq.~\eqref{eq:czgate}). The dotted line encapsulates CVCS construction. After measuring the top qumode with outcome $m_{1}$, the following operations are applied to the output $\ket{\psi}$: logic gate $\hat{U}_{1}$ (see Eq.~\eqref{eq:mbqcunitary}), a displacement, $\hat{X}(\frac{m_{1}}{g})$ (see Eq.~\eqref{eq:cvdisplacements}), and noise $\hat{N}(\varepsilon$) (see Eq.~\eqref{eq:cvnoiseoperator}). $\mathbf{(b)}$ Graphical representation of a CVW with an input state on the leftmost-qumode. When using the graphical representation, we must assume that the input state is Gaussian~\cite{gcfgps}. In Sec.~\ref{subsec:wignersingle} we use Wigner functions to generalize the description to mixed input states. The blue solid part of the wire is represented in the circuit diagram above. }\label{fig:circuitmbqc}
\end{figure}
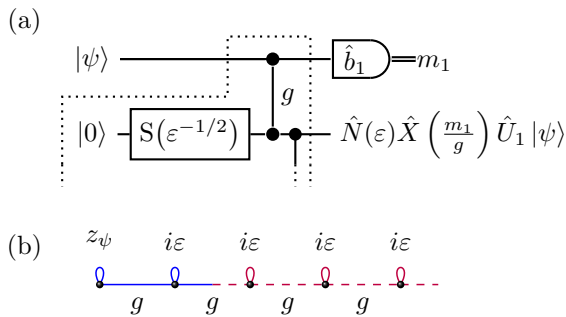

Now let us return to characterizing CVW measurements. It is sufficient to consider the measurement of a small portion of the CVW, as illustrated in Fig.~\ref{fig:circuitmbqc} for one measurement with a single-qumode input state $\ket{\psi}$.\footnote{If this is the beginning of a computation, then we can set $\ket{\psi_{\text{in}}}=\hat{S}(s)\ket{0}$  (with $s \gg 1$). Otherwise we can assume $\ket{\psi_{\text{in}}}$ is just the output of some previous computation step on the cluster or a state injected onto the cluster by an entangling operation with the left-most node.\label{ftnote:input}} The input state is encoded on the left-most wire node, which we label as the $i^{\text{th}}$ node. In the Schr\"odinger picture, the measurement of the $i^{\text{th}}$ node translates the input state $\ket{\psi}_{i}$ one node to the right and applies the following operation on the encoded input state:
\begin{equation}
\ket{\psi}_{i}\mapsto\hat{N}(\varepsilon ) \hat{X} \left( \frac{m_{i}}{g} \right)  \hat{U}_{i}\ket{\psi}_{i+1}, \label{eq:CVWops}
\end{equation}
where we refer to $\hat{U}_{i}$ as the logic gate (after one measurement), $\hat{X} ( \frac{m_{i}}{g} )$ as the displacement, and $\hat{N}(\varepsilon )$ as the noise operator. These are discussed below. 

The logic gate $\hat{U}_{i}$ can be decomposed as
\begin{align}
\hat{U}_{i}= \hat{F} \hat{S} (g) \hat{P} (\sigma_{i} ),\label{eq:mbqcunitary}
\end{align}
which are all defined above. Any single-qumode Gaussian unitary can be decomposed into a finite sequence of $\hat{U}_{i}$'s (up to displacements)~\cite{ulbttowqc}. 
While such gates are parameterized by three degrees of freedom in general, at least four CVW measurements are required in order to implement them \cite{ulbttowqc}. 
In principle one might naively think that with three measurement degrees of freedom, it is at least possible to get arbitrarily close to all single-qumode Gaussian unitaries. We shall see later in Sec.~\ref{subsec:3vs4measure} that this is ruled out in practice because the noise from finite squeezing diverges around the unachievable gates (independent of the amount of squeezing). 
For this reason, we assume four measurements are used, as this avoids such divergences while still being sufficient for implementing arbitrary single-qumode Gaussian unitaries~\cite{eogcc, ulbttowqc}.  

Next we define the phase-space displacements,
\begin{equation} \label{eq:cvdisplacements}
 \hat{X}(u) \coloneqq \exp\left(  -i u \hat{p}\right) , \hspace{1cm} \hat{Z}(v) \coloneqq \exp\left( i v \hat{q} \right).
\end{equation}
In the Heisenberg picture 
\begin{equation}
\hat{q} \xmapsto[]{\hat{X}(u )}  \hat{q} + u,  \hspace{1.75 cm} \hat{p} \xmapsto[]{\hat{Z}(v )}  \hat{p} + v.
\end{equation}
In either case, the other quadrature is left alone. In general, the position-quadrature ($\hat{q}$) displacement operator $\hat{X}(\frac{m}{g})$ which acts on the the output state in Eq.~\eqref{eq:CVWops} has to be corrected for, either by applying the inverse operation on the output, which we call the \emph{correction}
\begin{equation}
\hat{C}\coloneqq\hat{X} \left(-\frac{m}{g}\right) \label{eq:correction},
\end{equation}
or by adaptive measurement protocols, with future measurements depending on prior measurement outcomes. One caveat is that for measurement-based implementations of Gaussian unitaries, the adaptive measurement protocol is trivial: future measurements do not need to be adjusted based on prior measurement outcomes. Thus, only a final phase-space displacement correction is required for these protocols~\cite{qcwcvc}. 

The noise operator from Eq.~\eqref{eq:CVWops} is
\begin{equation} \label{eq:cvnoiseoperator}
\hat{N} \left( \varepsilon \right) \propto\exp \left( \frac{-\varepsilon\hat{q}^{2}}{2 } \right),
\end{equation}
which is not unitary and requires the output state to be normalized afterward (hence the $\propto$ symbol). It applies noise from finite squeezing to the state. For large squeezing, $\varepsilon$ is small. In the Schr\"odinger picture, this operator multiplies the state's position-space wavefunction by a 0-mean Gaussian with variance~$\varepsilon^{-1}$, called a Gaussian envelope (usually large). Equivalently, it convolves the state's momentum-space wavefunction by a 0-mean Gaussian with variance~$\varepsilon$ (usually small). In terms of wavefunctions, these two actions are equivalent, and only one or the other is ever needed to describe the action of this operator. An intuitive explanation is that in the position-space representation, the part of the wavefunction corresponding to large displacements in position (from the origin) is suppressed by the envelope. In the momentum-space representation, the wavefunction is ``blurred out'' by the convolution~\cite{uqcwcvcs}.

As will be shown in Sec.~\ref{subsec:wignersingle}, in the Wigner-function representation this operation has two simultaneous actions: (1)~multiplying the state's Wigner function by a 0-mean Gaussian envelope in position with variance~$\tfrac 1 2 \varepsilon^{-1}$ and also (2)~convolving the state's Wigner function in momentum by a 0-mean Gaussian with variance~$\tfrac 1 2 \varepsilon$. 

Note that in terms of wavefunctions, just one action (either envelope or convolution) is needed to represent the complete action of this operator, while two actions are involved in the Wigner representation. Also note that the variance of the envelope and that of the convolution are both reduced by a factor of~2 when moving from the wavefunction representation to the Wigner representation. This can be understood as accounting for the fact that the wavefunction is an amplitude, while the Wigner function is a \text{(quasi-)probability}~\cite{mtqsol}.


Finally, notice that as $\varepsilon\rightarrow 0$, $\hat{N}(\varepsilon)\rightarrow \hat{I}$, where $\hat{I}$ is the identity. Thus, wires with small $\varepsilon$ introduce less noise, which accords with our intuition about large squeezing corresponding to a better-quality CVCS~\cite{uqcwcvcs,qcwcvc}. In the next section, we will investigate how the noise depends on the wire weight $g$ by considering gates implemented by multiple measurements.

\subsection{Noise dependence on $g$}\label{sec:remodelwire}
First, consider a wire with uniform weight ${g=1}$. Although the $\hat{N}(\varepsilon)$ operator applied after one measurement introduces noise into the $\hat{p}$ quadrature, the Fourier transform that also gets applied will cause the noise from subsequent measurements to be distributed across both quadratures in a manner that depends also on the shearing parameters. In the case when all the shearing parameters are set to zero, noise will be added equally between the quadratures and the gate applied after $n$ measurements will be $\hat{F}^{n}$ (note that $\hat{F}^{4}=\hat{I}$). 

For $g\neq 1$, there is an additional squeezing operation that rescales the relative weight of some of the noise (see Eq.~\eqref{eq:mbqcunitary}) assuming that the ideal logical gate ($\hat{U}_{n}\dotsm \hat{U}_{2} \hat{U}_{1}$) is fixed. This is easy to see by considering an even number of measurements and ignoring the displacement terms, resulting in the total operation shown below:
\begin{multline}
\hat{N}(\varepsilon ) \hat{F}\hat{S}(g)\hat{P}(\sigma_{n} )\hat{N}(\varepsilon ) \hat{F}\hat{S}(g)\hat{P}(\sigma_{n-1} )\dotsm  \\ \dotsm\hat{N}(\varepsilon ) \hat{F}\hat{S}(g)\hat{P}(\sigma_{2} )\hat{N}(\varepsilon ) \hat{F}\hat{S}(g)\hat{P}(\sigma_{1} ).
\end{multline}
Commuting every odd squeezing operation to the left, past the following Fourier transform, noise, and shearing operations, we get
\begin{multline}
\hat{N}(\varepsilon ) \hat{F}\hat{P}(\sigma_{n}^{\prime})\hat{N}(\varepsilon g^{-2} ) \hat{F}\hat{P}(\sigma_{n-1} )\dotsm  \\ \dotsm \hat{N}(\varepsilon ) \hat{F}\hat{P}(\sigma_{2}^{\prime} ) \hat{N}(\varepsilon g^{-2}) \hat{F}\hat{P}(\sigma_{1} ). \label{eq:remodelwire1}
\end{multline}
Where $\sigma_{i}^{\prime}=\sigma_{i} g^{-2}$ for $i$ even. The form of the above expression (Eq.~\eqref{eq:remodelwire1}) has the equivalent interpretation of the operation that is applied when a $g=1$ wire is measured, only with every second noise parameter being rescaled. We can realise this graphically if we interpret this rescaling instead as the application of a local squeezing operation $\hat{S}(g)$ on every second wire node. Then, assuming a Gaussian input state labeled by the complex self-loop weight $z_{\psi}$, we have have effectively remodeled the weight-$g$ wire 
\begin{equation}
\hspace{0.5cm}
%
\begin{tikzpicture}[scale=1]
\path (0,0) coordinate (origin);
\path (1,0) coordinate (P1);
\path (2,0) coordinate (P3);
\path (3,0) coordinate (P5);
\path (4, 0) coordinate (E3);
\path (0.5, 0) node[label={[shift={(0,-0.65)}]$g$}] {};
\path (1.5, 0) node[label={[shift={(0,-0.65)}]$g$}] {};
\path (2.5, 0) node[label={[shift={(0,-0.65)}]$g$}] {};
\path (3.5, 0) node[label={[shift={(0,-0.65)}]$g$}] {};
\path (P3) edge[ out=120, in=60, looseness=1, loop, distance=0.35cm, color = blue, line width=0.2mm]
            node[above=3pt] {
$\color{black}i\varepsilon\color{blue}$ 
} (P3);

\path (P5) edge[ out=120, in=60, looseness=1, loop, distance=0.35cm, color = blue, line width=0.2mm]
            node[above=3pt] {
$\color{black}i\varepsilon\color{blue}$
} (P5);

\path (P1) edge[ out=120, in=60, looseness=1, loop, distance=0.35cm, color = blue, line width=0.2mm]
            node[above=3pt] {
$\color{black}i\varepsilon\color{blue}$
} (P1);

\path (origin) edge[ out=120, in=60, looseness=1, loop, distance=0.35cm, color = blue, line width=0.2mm]
            node[above=3pt] {
$\color{black}z_{\psi}\color{blue}$
} (origin);
\path (E3) edge[ out=120, in=60, looseness=1, loop, distance=0.35cm, color = blue, line width=0.2mm]
            node[above=3pt] {$\color{black}i\varepsilon\color{blue}$} (E3);
 \begin{scope}[color=blue,line width=0.2mm]
\draw (origin) -- (P1) (P1) -- (P3) (P3) -- (P5) (P5) -- (E3);
\end{scope}
 \begin{scope}[color=blue,line width=0.2mm, dashed]
\draw  (E3)-- (4.5, 0);
\end{scope}
 \shade[ball color=black] (0,0) circle(0.05);
 \shade[ball color=black] (1, 0) circle(0.05);
  \shade[ball color=black] (2,0) circle(0.05);
\shade[ball color=black] (3,0) circle(0.05);
\shade[ball color=black] (4,0) circle(0.05);
\end{tikzpicture}
\end{equation}
into a weight-1 wire with non-uniform self-loop weights: 
\begin{equation}
\hspace{0.5cm}
%
\begin{tikzpicture}[scale=1]
\path (0,0) coordinate (origin);
\path (1,0) coordinate (P1);
\path (2,0) coordinate (P3);
\path (3,0) coordinate (P5);
\path (4, 0) coordinate (E3);
\path (0.5, 0) node[label={[shift={(0,-0.65)}]$1$}] {};
\path (1.5, 0) node[label={[shift={(0,-0.65)}]$1$}] {};
\path (2.5, 0) node[label={[shift={(0,-0.65)}]$1$}] {};
\path (3.5, 0) node[label={[shift={(0,-0.65)}]$1$}] {};
\path (P3) edge[ out=120, in=60, looseness=1, loop, distance=0.35cm, color = blue, line width=0.2mm]
            node[above=3pt] {
$\color{black}i\varepsilon\color{blue}$ 
} (P3);

\path (P5) edge[ out=120, in=60, looseness=1, loop, distance=0.35cm, color = blue, line width=0.2mm]
            node[above=3pt] {
$\color{black}i\varepsilon g^{-2}\color{blue}$
} (P5);
\path (P1) edge[ out=120, in=60, looseness=1, loop, distance=0.35cm, color = blue, line width=0.2mm]
            node[above=3pt] {
$\color{black}i\varepsilon g^{-2}\color{blue}$
} (P1);
\path (origin) edge[ out=120, in=60, looseness=1, loop, distance=0.35cm, color = blue, line width=0.2mm]
            node[above=3pt] {
$\color{black}z_{\psi} \color{blue}$
} (origin);
\path (E3) edge[ out=120, in=60, looseness=1, loop, distance=0.35cm, color = blue, line width=0.2mm]
            node[above=3pt] {$\color{black}i\varepsilon \color{blue}$} (E3);
\begin{scope}[color=blue,line width=0.2mm]
\draw (origin) -- (P1) (P1) -- (P3) (P3) -- (P5) (P5) -- (E3);
\end{scope}
 \begin{scope}[color=blue,line width=0.2mm, dashed]
\draw (E3)-- (4.5, 0);
\end{scope}
 \shade[ball color=black] (0,0) circle(0.05);
 \shade[ball color=black] (1, 0) circle(0.05);
  \shade[ball color=black] (2,0) circle(0.05);
\shade[ball color=black] (3,0) circle(0.05);
\shade[ball color=black] (4,0) circle(0.05);
\end{tikzpicture}
\label{eq:rescalewire1}
\end{equation}
Note that we have assumed that the input state is Gaussian in order to represent the CVW state using the graphical calculus~\cite{gcfgps}. This provides an intuitive pictorial representation of the remodeling procedure. The result is fully general, however, and applies to arbitrary inputs, including mixed states.\footnote{This can be verified straightforwardly with Wigner functions using the methods employed in Sec.~\ref{subsec:wignersingle}. We leave an explicit proof as an exercise for the reader.} 

The above shows that it is possible to encorporate the change in wire weight ($1\mapsto g$) into a rescaling by $g^{-2}$ of the measurement basis and noise parameter, on half the nodes. Consequently, the lower the value of $g$, the higher the overall noise introduced. In fact, the parameter $g$ sets the noise bias between the quadratures since the noise alternates between the quadratures due to the Fourier transform, and only the $\hat{q}$-quadrature noise gets rescaled. In terms of overall noise, lower-weight wires amplify the noise from finite squeezing and are therefore suboptimal. 

Another way to take the weight $g$ into account is to rescale the measurement outcomes of every node by $\sqrt{g}$, like applying a $\hat{S}\left(\sqrt{g}\right)$ on every node as shown on the wire below.
\begin{equation}
\hspace{0.25cm}
%
\begin{tikzpicture}[scale=1]
\path (0,0) coordinate (origin);
\path (1,0) coordinate (P1);
\path (2,0) coordinate (P3);
\path (3,0) coordinate (P5);
\path (4, 0) coordinate (E3);
\path (0.5, 0) node[label={[shift={(0,-0.65)}]$1$}] {};
\path (1.5, 0) node[label={[shift={(0,-0.65)}]$1$}] {};
\path (2.5, 0) node[label={[shift={(0,-0.65)}]$1$}] {};
\path (3.5, 0) node[label={[shift={(0,-0.65)}]$1$}] {};
\path (P3) edge[ out=120, in=60, looseness=1, loop, distance=0.35cm, color = blue, line width=0.2mm]
            node[above=3pt] {
$\color{black}i\varepsilon g^{-1}\color{blue}$ 
} (P3);
\path (P5) edge[ out=120, in=60, looseness=1, loop, distance=0.35cm, color = blue, line width=0.2mm]
            node[above=3pt] {
$\color{black}i\varepsilon g^{-1}\color{blue}$
} (P5);
\path (P1) edge[ out=120, in=60, looseness=1, loop, distance=0.35cm, color = blue, line width=0.2mm]
            node[above=3pt] {
$\color{black}i\varepsilon g^{-1}\color{blue}$
} (P1);
\path (origin) edge[ out=120, in=60, looseness=1, loop, distance=0.35cm, color = blue, line width=0.2mm]
            node[above=3pt] {
$\color{black}z_{\psi} g^{-1}\color{blue}$
} (origin);
\path (E3) edge[ out=120, in=60, looseness=1, loop, distance=0.35cm, color = blue, line width=0.2mm]
            node[above=3pt] {$\color{black}i\varepsilon g^{-1}\color{blue}$} (E3);
 \begin{scope}[color=blue,line width=0.2mm]
\draw (origin) -- (P1) (P1) -- (P3) (P3) -- (P5) (P5) -- (E3);
\end{scope}
 \begin{scope}[color=blue,line width=0.2mm, dashed]
\draw  (E3)-- (4.5, 0);
\end{scope}
 \shade[ball color=black] (0,0) circle(0.05);
 \shade[ball color=black] (1, 0) circle(0.05);
  \shade[ball color=black] (2,0) circle(0.05);
\shade[ball color=black] (3,0) circle(0.05);
\shade[ball color=black] (4,0) circle(0.05);
\end{tikzpicture}
\label{eq:remodel2wire}
\end{equation}
 This results in a CVW weighted uniformly along links and self-loops, with noise seemingly introduced in equal quantities to each quadrature (the self-loop weights of every odd node is the same as every even node). In doing so, the input state $z_{\psi}$ has effectively been squeezed on the wire ($z_{\psi}\mapsto z_{\psi} g^{-1} $). This can be thought of as an ``encoding" onto the effective ``$g=1$" wire with uniform self-loops ($\varepsilon^{\prime}=\varepsilon g^{-1}$). The measurement protocol must then be changed, $\hat{p}+\sigma_{i}\hat{q} \mapsto \hat{p}+ \sigma_{i} g^{-1} \hat{q}$, in order to effect an equivalent logic gate. The advantage is that the noise properties are described by a single parameter $\varepsilon^{\prime}$ up to the encoding. While it appears as though the noise is added to the quadratures in equal amounts, this is only after applying the encoding, which effectively rescales the quadratures relative to one another (and biases the un-biased noise structure). However, if the computational protocol is assumed to start with a blank ancilla squeezed by $\sqrt{\varepsilon g^{-1}}$ (rather than $\sqrt{\varepsilon}$, see footnote \ref{ftnote:input}), then we \emph{can} treat our wire as having weight $g=1$ with rescaled self-loops. 

Let us consider the types of CVWs generated from the DRW. By performing $\hat{q}$ measurements on all the top nodes, the DRW can be converted to the CVW with uniform edge weight $g_{\text{D}}$ and uniform self-loop weight $\varepsilon_{\text{D}}$ (see Eqs.~\eqref{eq:gdual} and \eqref{eq:edual}). The edge weight $g_{\text{D}}$ is upper bounded by the value $\tfrac{1}{2}$. By applying the parameter rescaling corresponding to Eq.~\eqref{eq:remodel2wire} and defining $\varepsilon^{\prime}_{D} \coloneqq \varepsilon_{\text{D}} g_{\text{D}}^{-1}$, we see that this protocol amplifies the effect of the self-loop noise that gets added: 
\begin{equation}
\hat{N}\left(\varepsilon_{\text{D}}^{\prime} \right) = \hat{N}\left( 2 \csch{2 \alpha} \right)  \to \hat{N}\left( 2 \varepsilon_{\text{D}} \right) \label{eq:noiseremodel}
\end{equation}
where the limit is that of large squeezing (${\alpha\rightarrow \infty}$). 
The arguments above highlight the importance of the CVW edge weight $g$ on the computation, demonstrating why using the CVW protocol is suboptimal: $g_{\text{D}}$ is small relative to $g=1$ for the standard CVW~\cite{uqcwcvcs, qcwcvc, eogcc}, and this results in more noise from finite squeezing. 

The purpose of the next section will be to derive a more quantitative description of the noise using the Wigner-function formalism. This will address the following issues: First, we wish to compare the CVW and macronode protocols in terms of how much noise is applied per gate, yet the noise will depend on the measurement bases, which will be competely different and could even involve a different number of measurements.  Second, when applying the measurement-dependent correction, the Gaussian envelope that is the manifestation of the noise from finite squeezing acting on the position-space wavefunction will be displaced relative to the origin. Thus, the noise depends upon the measurement outcome~\cite{qcwcvc}. The Wigner-function formalism allows us to describe mixed states, particularly the output states of MBQC that are averaged over possible measurement outcomes. This will help us to define a quantity that captures the average noise introduced by $n$ measurements and hence compare the measurement protocols in a \emph{noise-per-gate} manner.

\subsection{Wigner function formalism} \label{subsec:wignersingle}
Wigner functions are quasi-probability distributions that provide a phase-space picture for arbitrary mixed states. Given an arbitrary input state $\rho$, its Wigner function is given by~\cite{mtqsol}
\begin{equation}
W_{\hat \rho}( q, p) \coloneqq \frac{1}{2\pi} \int \mathrm{d} x \left\langle q-\frac{x}{2} \right|_{q} \hat{\rho} \left| q+\frac{x}{2}\right\rangle_{q} e^{ixp},
\end{equation} 
where the subscript $q$ labels position basis states, and $\mathbf{x}=\left( q_{1}, q_{2}, \dots p_{1}, p_{2}, \dots \right)^{T}$ is a vector of $c$-numbers.
We let $\mathbf{x}_{i}=\left( q_{i} , p_{i} \right)^{T}$ denote the $i^{\text{th}}$ qumode register.
 
We now define the action of unitary operations on quantum states in the Wigner-function formalism. Given a quantum state~$\hat \rho$ with Wigner function~$W_{\hat \rho}(q,p)$ as above, the state evolves under a unitary operator~$\hat B$ as $\hat \rho \mapsto \hat B \hat \rho \hat B^\dag$, whose Wigner function is $W_{\hat B \hat \rho \hat B^\dag}(q,p)$. We label the Wigner representation of an arbitrary unitary operator~$\hat B$ by the same symbol but with calligraphic font:
\begin{align}
	\mathcal{B}[W_{\hat\rho}(q,p)] \coloneqq W_{\hat B \hat \rho \hat B^\dag}(q,p),
\end{align}
for all unitaries~$\hat B$.

Under Gaussian unitary evolution $\hat{E}$ with Heisenberg-picture symplectic representation $\mathbf{E}$ defined by 
\begin{equation}
\hat{\mathbf x} \xmapsto{\;\;\hat E\;\;} \hat{E}^{\dagger}\hat{\mathbf{x}}\hat{E} \eqqcolon \mathbf{E}\mathbf{\hat{x}}+ \mathbf{c}, \label{eq:Edef}
\end{equation}
where $\hat{\mathbf{x}}=\left( \hat{q}_{1}, \hat{q}_{2}, \dots \hat{p}_{1}, \hat{p}_{2}, \dots \right)^{T}$, the Wigner-function arguments (which are $c$-numbers) update in the reverse way to the Heisenberg evolution of operators:
\begin{align} 
	\mathcal{E}[W_{\hat \rho}(\mathbf x)] &= W_{\hat E \hat \rho \hat E^\dag}(\mathbf{x}) \nonumber \\
	&= W_{\hat \rho}\bigl(\mathbf{E}^{-1}(\mathbf{x}-\mathbf{c})\bigr).\label{eq:wignerevolve}
\end{align}
Note that under the action of $\mathcal{E}$, the argument of the Wigner function updates using~$\mathbf E^{-1}$ after a displacement by $-\mathbf{c}$, while the Heisenberg evolution of quadrature operators due to $\hat{E}$ uses~$\mathbf E$ before a displacement by $+\mathbf{c}$ (Eq.~\ref{eq:Edef}).

Now we recast the evolution shown in Fig.~\ref{fig:circuitmbqc} in the Wigner formalism, extending the results in Ref.~\cite{qcwcvc} to weight-$g$ wires and single-qumode Gaussian unitaries. 
Define the Gaussian function $G_{y}(x)$ to be
\begin{equation}
G_{y}(x)= \frac {1}{\sqrt{\pi y}} \exp \left(-\frac {x^2}{y} \right),
\end{equation}
which is a normalized Gaussian with mean~0 and variance $y/2$. Incidentally, we can write the noise operator from Eq.~\eqref{eq:cvnoiseoperator} as $\hat{N} \left( \varepsilon \right) \propto G_{2 / \epsilon}(\hat q)$. 

Let the (possibly mixed) input state~$\rho_{\text{in}}$ be represented by the Wigner function~$W_{\text{in}}(q,p)$. A blank cluster qumode---i.e.,~a momentum-squeezed state with (large) squeezing factor $\varepsilon^{-1/2}$---is represented by the Wigner function~$G_{1/\varepsilon} (q) G_{\varepsilon} (p)$. The initial two-qumode state in Fig.~\ref{fig:circuitmbqc}, which consists of the input state attached via $\hat C_Z(g)$ to a blank cluster qumode, is represented by the following Wigner function:
\begin{equation} \label{eq:singlewignerin}
W_{\text{in}}\left( q_{1}, p_{1}-g q_{2} \right) G_{1/\varepsilon} (q_{2}) G_{\varepsilon} (p_{2}-g q_{1}).
\end{equation}

Define the symplectic-matrix representation of $\hat{U}_{i}$ (see Eq.~\eqref{eq:mbqcunitary}) to be 
\begin{equation}
\mathbf{\mathbf{U}}_{i}=\begin{pmatrix} -\sigma_{i} g^{-1} & -g^{-1} \\ g & 0 \end{pmatrix}.\label{eq:Usymrep}
\end{equation}
Then the Wigner function~$W_{\text{out}}(\mathbf{x}_{2})$ for the output state after a single CVW node measurement (in the basis $\hat{p}+\sigma_{1} \hat{q}$) and after applying the correction operator $\hat{C}$ (see Eq.~\eqref{eq:correction}) is given by 
\begin{align}\hspace{-0.8cm}
P(m_{1}) W_{\text{out}} (\mathbf{x}_{2} ) &= \int \! \mathrm{d} \tau_{2}  W_{\text{in}} \left( \mathbf{\mathbf{U}}^{-1}_{1}\left( \mathbf{x}_{2}+g \boldsymbol\tau \right) \right) \nonumber \\ & \times G_{\varepsilon} \left( g \tau_{2} \right) G_{1/\varepsilon} \left(q_{2} + \frac{m_{1}}{g}\right),   \label{outafter1}
\end{align}
where $P(m_{1})$ is the probability of measuring outcome $m_{1}$, and $\boldsymbol\tau=\left( 0 , \tau_{2} \right)^{T}$. 
Eq.~\eqref{outafter1} shows that the noise from finite squeezing manifests as both a phase-space Gaussian convolution in momentum and a Gaussian envelope in position with measurement-outcome-dependent mean. As the measurement outcomes will be different each time, we consider the measurement-averaged distortion on the output state:
\begin{align} \hspace{-0.8cm}
W_{\text{avg}} (\mathbf{x}_{2}) & = \int \! \mathrm{d} m_{1} P(m_{1}) W_{\text{out}} (\mathbf{x}_{2}) \nonumber \\ & =\int \! \mathrm{d} \tau^{\prime}_{2}  W_{\text{in}} \left( \mathbf{\mathbf{U}}_{1}^{-1}\left( \mathbf{x}_{2}+ \boldsymbol\tau^{\prime} \right) \right) G_{\varepsilon} \left(  \tau_{2}^{\prime} \right),
\label{eq:cvwirewigneroutput} 
\end{align}
where $\boldsymbol\tau^{\prime}= g \boldsymbol\tau=\left( 0 , \tau^{\prime}_{2} \right)^{T}=\left( 0 , g\tau_{2} \right)^{T}$. Thus, the average effect of noise from finite squeezing is a Gaussian convolution, similar to the $g=1$ case treated in Ref.~\cite{qcwcvc}.
\blk

Iterating the above expression yields the average Wigner function $W_{\text{avg}}^{(n)} \left(\mathbf{x}\right)$ after $n$ homodyne measurements $\hat{b}_{i}= \hat{p} + \sigma_{i} \hat{q}$ on an $n$-node CVW. Define $W_{\text{avg}}^{(0)}(\mathbf{x}) \coloneqq W_{\text{in}}(\mathbf{x})$. 
Then, 
\begin{align} \hspace{-0.8cm}
W_{\text{avg}}^{(n)} \left( \mathbf{x}_{n} \right) &= \int \! \mathrm{d} \tau_{2}^{(n)} W_{\text{avg}}^{(n-1)} \left( \mathbf{U}_{n}^{-1} ( \mathbf{x}_{n} + \boldsymbol\tau^{(n)} ) \right) \nonumber \\ &\quad \times G_{\varepsilon} \left( \tau^{(n)}_{2} \right),  \label{eq:singlewigneriterative}
\end{align}
where $\boldsymbol\tau^{(n)}= ( 0, \tau_{2}^{(n)} )^{T}$, and $\mathbf{U}_{n}$ is the symplectic matrix representation of the Heisenberg action of $\hat{U}_{n}$. Just as in Eq.~\eqref{eq:cvnoiseoperator}, each $i^{\text{th}}$ measurement convolves $W_{\text{avg}}^{(i-1)}$ in the momentum quadrature. Now we have expressed the average output state using the Wigner formalism. 

We are interested in characterizing MBQC on the CVW in terms of how much noise is added from finite squeezing. There are a couple of different ways to ``unpack" Eq.~\eqref{eq:singlewigneriterative} into a description involving the desired computation and the added noise. If the noise and displacements are ignored, then measuring the first $n$ nodes on a CVW applies the operation 
\begin{equation}
\hat{U}_{\boldsymbol{\sigma}}=\hat{U}_{n}\dotsm\hat{U}_{2}\hat{U}_{1},
\end{equation}
where the overall unitary applied depends on the shearing parameters $\boldsymbol\sigma\coloneqq (\sigma_{1},\dots ,\sigma_{n})$. This can be thought of as the ideal operation applied in the absence of noise and after the displacements are corrected for. Define
\begin{equation}
\tilde{\mathbf{U}}_{i}\coloneqq \mathbf{U}_{i}\dotsm\mathbf{U}_{1},
\end{equation}
and
\begin{equation}
\check{\mathbf{U}}_{i}\coloneqq\mathbf{U}_{n}\dotsm\mathbf{U}_{i+1},
\end{equation}
where $n$ is the total number of homodyne measurements made on the CVW. Note that ${\check{\mathbf{U}}_{0} = \check{\mathbf{U}}_{i} \tilde{\mathbf{U}}_{i} = \tilde{\mathbf{U}}_{n}}$, which is the symplectic representation of the Heisenberg action of the full~$\hat U_{\boldsymbol\sigma}$. The Wigner function for the (ideal) output state without noise and displacements (i.e.,~ignoring $\hat{N}(\varepsilon) \hat{X}(\tfrac {m_i}{g})$ in Eq.~\eqref{eq:CVWops}) is 
\begin{equation}
W_{\text{ideal}}(\mathbf{x})\coloneqq W_{\text{in}}\left( \tilde{\mathbf{U}}^{-1}_{n}\mathbf{x} \right).
\end{equation}
Using this, Eq.~\eqref{eq:singlewigneriterative} can be expanded out as
\begin{align}
\hspace{-0.8cm}
W^{(n)}_{\text{avg}} \left( \mathbf{x}_{n} \right) &=  \int \mathrm{d} \tau^{(n)}_{2}\dotsm \mathrm{d} \tau^{(1)}_{2} W_{\text{ideal}} \left( \mathbf{x}_{n} + \sum^{n}_{i=1} \check{\mathbf{U}}_{i} \boldsymbol\tau^{(i)} \right) \nonumber \\ &\quad\times G_{\varepsilon} \left( \tau^{(n)}_{2} \right)\dotsm  G_{\varepsilon} \left( \tau^{(1)}_{2} \right). \label{eq:navgwigner}
\end{align}
This expression shows us that the average effect of $n$ homodyne measurements relative to the ideal unitary evolution (which maps $W_{\text{in}}\mapsto W_{\text{ideal}}$) is $n$ gate-dependent Gaussian convolutions. Let $\check{\mathcal{N}}_{\boldsymbol\sigma}$ denote the application of these Gaussian convolutions to an arbitrary Wigner function~$W(\mathbf{x})$:
\begin{align} \hspace{-0.8cm}
\check{\mathcal{N}}_{\boldsymbol\sigma}[W\left( \mathbf{x} \right)] &\coloneqq \! \int \! \mathrm{d} \tau^{(n)}_{2}\!\! \dotsm\mathrm{d} \tau^{(1)}_{2} W\!\! \left( \mathbf{x} \! + \! \sum^{n}_{i=1} \check{\mathbf{U}}_{i} \boldsymbol\tau^{(i)} \right) \nonumber \\ &\quad\times G_{\varepsilon} \left( \tau^{(n)}_{2} \right)\dotsm  G_{\varepsilon} \left( \tau^{(1)}_{2} \right). 
\end{align}
The Wigner function $W^{(n)}_{\text{avg}}$ can then be expressed as
\begin{align}
W^{(n)}_{\text{avg}}(\mathbf{x}_{n}) &= \check{\mathcal{N}}_{\boldsymbol\sigma}\circ\mathcal{U}_{\boldsymbol\sigma}[W_{\text{in}}(\mathbf{x}_{n})] \nonumber \\ &= \check{\mathcal{N}}_{\boldsymbol\sigma}[W_{\text{ideal}}(\mathbf{x}_{n})]. 
\end{align}
Thus, $\check{\mathcal{N}}_{\boldsymbol\sigma}$ can be thought of as the average noise added by the $n$-measurement channel \emph{after} a perfect logic gate $\hat{U}_{\boldsymbol\sigma}$ is applied. 

To arrive at an expression for the effect of the noise from finite squeezing \emph{before} the perfect logic gate is applied, consider \emph{undoing} the ideal unitary---i.e.,~applying $\hat{U}_{\boldsymbol\sigma}^{\dagger}$---to see how the noise causes the output Wigner function to differ from that of the input. By using Eq.~\eqref{eq:wignerevolve}, we can define a new Wigner function for this case:
\begin{align}
W^{(n)}_{\text{undo}} (\mathbf{x}_{n}) &\coloneqq \mathcal{U}^{-1}_{\boldsymbol\sigma}\left[ W_{\text{avg}}^{(n)}(\mathbf{x}_{n}) \right] \nonumber\\ &= W_{\text{avg}}^{(n)}(\tilde{\mathbf{U}}_{n}\mathbf{x}_{n}). 
\end{align}
This is the average Wigner function after performing $n$ measurements, applying the correction operators, and undoing the ideal logic gate~$\hat{U}_{\boldsymbol\sigma}$. 
 Expanding this gives 
\begin{align}\hspace{-0.8cm}
W^{(n)}_{\text{undo}}\! \! \left( \mathbf{x}_{n} \right) &=\! \int \! \mathrm{d} \tau^{(n)}_{2}\!\! \dotsm\mathrm{d} \tau^{(1)}_{2} W_{\text{in}} \left( \mathbf{x}_{n} \! + \! \sum^{n}_{i=1} \tilde{\mathbf{U}}_{i} \boldsymbol\tau^{(i)} \right) \nonumber \\ &\quad\times G_{\varepsilon} \left( \tau^{(n)}_{2} \right)\dotsm  G_{\varepsilon} \left( \tau^{(1)}_{2} \right). \label{eq:nundowigner}
\end{align}
Similar to the above, we define $\tilde{\mathcal{N}}_{\boldsymbol\sigma}$ to be the map that applies these Gaussian convolutions to an arbitrary Wigner function~$W (\mathbf{x})$:
\begin{align} \hspace{-0.8cm}
\tilde{\mathcal{N}}_{\boldsymbol\sigma}[W ( \mathbf{x} )] &\coloneqq\! \int \! \mathrm{d} \tau^{(n)}_{2} \!\! \dotsm\mathrm{d} \tau^{(1)}_{2} W\!\! \left( \mathbf{x} \! + \! \sum^{n}_{i=1} \tilde{\mathbf{U}}_{i} \boldsymbol\tau^{(i)} \right) \nonumber \\ & \quad\times G_{\varepsilon} \left( \tau^{(n)}_{2} \right)\dotsm  G_{\varepsilon} \left( \tau^{(1)}_{2} \right).
\end{align}
Since $W^{(n)}_{\text{undo}} (\mathbf{x}_{n}) = \tilde{\mathcal{N}}_{\boldsymbol\sigma}[W_{\text{in}}(\mathbf{x}_{n})]$, we can expand the average output state as
\begin{align}
W_{\text{avg}}^{(n)} (\mathbf{x}_{n})&= \mathcal{U}_{\boldsymbol\sigma}\left[W^{(n)}_{\text{undo}} (\mathbf{x}_{n})\right] \nonumber\\ &=  \mathcal{U}_{\boldsymbol\sigma}\circ \tilde{\mathcal{N}}_{\boldsymbol\sigma}[W_{\text{in}}(\mathbf{x}_{n})].
\label{eq:noisebeforegatedecomp}
\end{align}
Thus, $\tilde{\mathcal{N}}_{\boldsymbol\sigma}$ can be thought of as the average noise added by the $n$-measurement channel \emph{before} a perfect logic gate $\hat{U}_{\boldsymbol\sigma}$ is applied. 

We have two equivalent descriptions for the output of the $n$-homodyne CVW channel, shown by Eqs.~\eqref{eq:navgwigner} and \eqref{eq:nundowigner}, which are quite similar in form. These expressions can be simplified somewhat. In what follows, we will restrict our attention to the noise-before-gate decomposition (Eq.~\eqref{eq:nundowigner}). The other case can be derived analogously. 

For ${n\geq 2}$ measurements, we can replace the $n$ convolutions---each along a single phase-space direction---with a single \emph{bivariate} phase-space convolution that will depend on all $n$ measurements, as shown below. This is guaranteed to have Gaussian form, as the convolution of $n$ Gaussians is itself Gaussian. Thus, 
\begin{equation} \label{eq:wignerafternsingle}
W^{(n)}_{\text{undo}} \left( \mathbf{x}_{n} \right) = \int \! \mathrm{d} \kappa_{1}  \mathrm{d} \kappa_{2} W_{\text{in}} \left( \mathbf{x}_{n}+\boldsymbol\kappa \right) B_{\mathbf{\Sigma}_{n}} (\boldsymbol \kappa ), 
\end{equation}
where $\boldsymbol\kappa=\left( \kappa_{1} , \kappa_{2} \right)^{T}$ contains the new dummy variables for the bivariate convolution, replacing $\boldsymbol\tau^{(1)},\boldsymbol\tau^{(2)},\dots ,\boldsymbol\tau^{(n)}$, and
\begin{equation}
B_{\mathbf{K}}(\boldsymbol\kappa )=\left( \pi \sqrt{\det(\mathbf{K}}) \right)^{-1} \exp(-\boldsymbol\kappa^{T} \mathbf{K}^{-1} \boldsymbol\kappa)\label{eq:bivariateconv}
\end{equation}
is a normalized bivariate Gaussian distribution with covariance matrix $\tfrac{1}{2}\mathbf{K}$. Then,
\begin{equation}
\mathbf{\Sigma}_{n}=\sum^{n}_{i=1}  \tilde{\mathbf{U}}_{i}^{-1} \mathbf{\Sigma}_{*} \tilde{\mathbf{U}}_{i}^{-T}\!\!\!\! ,  \label{eq:sigmansingle}
\end{equation}
where
\begin{equation}
\mathbf{\Sigma}_{*}\coloneqq\begin{pmatrix} 0 & 0 \\ 0 & \varepsilon \end{pmatrix}\hspace{-0.1cm}. \label{eq:sigmastarsingle}
\end{equation}
We can interpret $\mathbf{\Sigma}_{*}$ as the covariance matrix for a single Gaussian convolution that represents the action of the $\hat{N}(\varepsilon )$ operator from Eq.~\eqref{eq:cvnoiseoperator} averaged over measurement outcomes.
For future use in the Appendices, we also define 
\begin{equation}
\mathbf{\Sigma}^{*}\coloneqq\begin{pmatrix} \varepsilon & 0 \\ 0 & 0 \end{pmatrix}\hspace{-0.1cm}. \label{eq:sigmaupstarsingle}
\end{equation}

Recall we are using the noise-before-gate decomposition (Eq.~\eqref{eq:noisebeforegatedecomp}). The analogous expression to Eq.~\eqref{eq:wignerafternsingle} for the noise-after-gate decomposition (Eq.~\eqref{eq:navgwigner}) is 
\begin{equation}
W^{(n)}_{\text{avg}}(\mathbf{x}_{n})= \int \! \mathrm{d} \kappa_{1}  \mathrm{d} \kappa_{2} W_{\text{ideal}} \left( \mathbf{x}_{n}+\boldsymbol\kappa \right) B_{\mathbf{\boldsymbol\Sigma}^{\prime}_{n}} (\boldsymbol \kappa ), 
\end{equation}
where
\begin{equation}
\mathbf{\Sigma}^\prime_{n}=\sum^{n}_{i=1}  \check{\mathbf{U}}_{i}^{-1} \mathbf{\Sigma}_{*} \check{\mathbf{U}}_{i}^{-T}.  \label{eq:sigmansingle}
\end{equation}
For the rest of this Article, we will focus on the noise-before-gate description of CVW computation (as in Eq.~\eqref{eq:noisebeforegatedecomp}).

The covariance matrix $\mathbf{\Sigma}_{n}$ is the only part of equation Eq.~\eqref{eq:wignerafternsingle} that depends on the measurements in any way. It characterises the bivariate Gaussian convolution in terms of  the number of measurements, which observable is measured, and ultimately the gate that was performed. 

Consider the trace of $\tfrac{1}{2}\mathbf{\Sigma}_{n}$, which is invariant under phase-space rotations. It is the sum of the variances along any two orthogonal phase-space directions, such as $q$ and $p$. Alternatively, by changing from cartesian coordinates~$(q,p)$ to polar coordinates~$(r,\theta)$, with $r=\sqrt{q^{2}+p^{2}}$, this quantity can be interpreted as the average \emph{radial variance} of the Wigner function (averaging uniformly over~$\theta$) since 
$\Delta r^{2} = \Delta q^{2} + \Delta p^{2}$. 
Convolving $W_{\text{in}}$ with $B_{\mathbf{\Sigma}}$ adds the variance of the bivariate Gaussian distribution to that of the input state. Thus, we can quantify the noise added to the state by defining the \emph{scalar variance} 
\begin{equation}
SV(n) \coloneqq \tfrac{1}{2}\tr[\mathbf{\Sigma}_{n}],  \label{eq:scalarvariance}
\end{equation}
which quantifies the noise of the average output Wigner function, which in turn depends on the number of measurements, the choice of each measurement basis, and parameters $\varepsilon$ and $g$. 

As we shall see later, there are many ways of applying gates through single-qumode measurements on the CVW. In terms of the scalar variance (or how much noise is added), these methods will not be equivalent, and it will be useful to define the minimized scalar variance 
\begin{equation}
\mathcal{SV}(n) \coloneqq \min_{\{\boldsymbol\sigma | \tilde{\mathbf{U}}_{n}=\mathbf{E}\}}{SV(n)}, \label{eq:CVWminSV}
\end{equation}
 where the minimization is over the measurement degrees of freedom (i.e.,~shearing parameters/homodyne angles) with the constraint that the total unitary applied is equal to the desired gate unitary $\hat{E}$---i.e.,~${\tilde{\mathbf{U}}_{n}=\mathbf{E}}$. Throughout this Article, we will use calligraphic font to denote the minimized version of the scalar variance with respect to any free measurement degrees of freedom and some gate. This will allow for a fair comparison between the CVW and macronode protocols.

The $g$ dependence of $SV(n)$ can be expressed simply for an even number of measurements $n$. Assuming that the logic gate in the noiseless limit ($\hat{U}_{\boldsymbol\sigma}$) is fixed, applying a rescaling of the shearing parameters as in Eq.~\eqref{eq:rescalewire1} reduces the dependence of the scalar variance in terms of $g$ to be 
\begin{align}
SV(n)=\sum^{n/2}_{k=1} \left(f_{2k-1} + g^{-2} f_{2k} \right), \label{eq:scalardependenceg}
\end{align}
where each $f_{i}$ is a positive multivariate polynomial (defined below) in the shearing parameters $\sigma_{1}, \sigma^{\prime}_{2},\dots,\sigma_{i}^{(\prime)}$,  where the parameters with even indices have been rescaled, namely $\sigma_{j}^{\prime}=\sigma_{j} g^{-2}$, and all the shearing parameters are fixed by requiring that the measurement procedure effect the gate~$\hat{U}_{n}\dotsm\hat{U}_{2}\hat{U}_{1}$ and by the condition that $SV(n)$ is minimized. Then,
\begin{align}
f_{i}=\frac{1}{2}\tr{( \tilde{\mathbf{T}}_{i}^{-1} \mathbf{\Sigma}_{*} \tilde{\mathbf{T}}_{i}^{-T})},
\end{align}
where $\tilde{\mathbf{T}}_{i}=\mathbf{T}_{i}\mathbf{T}_{i-1}\dotsm\mathbf{T}_{1}$, with $\mathbf{T}_{j}=\mathbf{F} \mathbf{P}(\sigma_{j}^{(\prime)})$, where $\sigma_{j}$ is primed for even $j$'s only. $\mathbf{T}_{j}$ is just $\mathbf{U}_{j}$ after remodelling to a $g=1$ CVW (by rescaling the shearing parameters). Thus, we have that $\mathbf{T}_{j+1}\mathbf{T}_{j}=\mathbf{U}_{j+1}\mathbf{U}_{j}$, $\forall j\in \mathbb{N}$. Defining the $f_{i}$ in terms of rescaled shearing parameters and $\mathbf{T}_{i}$'s suppresses their explicit dependence on $g$.

 Eq.~\eqref{eq:scalardependenceg} shows that $SV(n)\rightarrow\infty$ as $g\rightarrow 0$ when the ideal gate (i.e., in the noiseless limit) is fixed. The $g\rightarrow 0$ limit can be understood as the ``unconnected cluster limit," where we expect no information to propagate along the wire. In the large-$g$ limit, only the even $f_{i}$ terms disappear, in analogy to Eq.~\eqref{eq:remodelwire1}. Also note that in the infinite-squeezing limit ($\varepsilon\rightarrow 0$), $SV(n) \rightarrow 0$. Thus, $B_{\mathbf{K}}(\boldsymbol\kappa ) \rightarrow \delta(\boldsymbol{\kappa})$ and hence,
\begin{equation}
\lim_{\varepsilon\rightarrow 0} W^{(n)}_{\text{undo}} \left( \mathbf{x}_{n} \right) = W^{(0)} (\mathbf{x}_{n}),
\end{equation}
as required.  
This analysis confirms what was discussed above, that CVWs with small $g$ weights amplify noise from finite squeezing. Motivated by this, we consider an alternative approach in the next section.

\section{Macronode protocol} \label{sec:dualrail}
Macronode-based computation does not involve converting the DRW into a CVW, and consequently, some features of the computation will differ due to use of the additional DRW structure. We can describe the each macronode by the vector of quadrature operators corresponding to its constituent physical qumodes
\begin{equation}
\mathbf{\hat{x}}_{i}=(\hat{q}_{ia}, \hat{q}_{ib}, \hat{p}_{ia}, \hat{p}_{ib})^T, 
\end{equation}
where $a$ and $b$ label distinct \emph{physical qumodes} comprising the macronode $i$. To treat the DRW as a double-thick quantum wire, we must define the single-qumode logical subspace within each macronode.

To this end, we define the quadrature operators
\begin{equation}
\hat{q}_{i\pm}\coloneqq \frac{1}{\sqrt{2}}\left(\hat{q}_{i a}\pm\hat{q}_{i b}\right), \quad \hat{p}_{i \pm}\coloneqq\frac{1}{\sqrt{2}}\left(\hat{p}_{i a}\pm\hat{p}_{i b}\right), \label{eq:distlabels}
\end{equation}
which correspond to the \emph{distributed modes} labeled $+$ and $-$. Note that the physical modes and distributed modes represent alternative tensor-product decompositions of the same two-qumode Hilbert space of macronode~$i$. This means that the entanglement structure of a given state will appear different depending on which tensor-product decomposition is used~\cite{qtpsaoi}. We label the physical modes as such because they correspond to the particular temporal modes~\cite{tmcvcsulo,ulscvcsmittd} or frequency modes~\cite{owqcitofc,wqofcicvhcs,eromeo60moaqofc} on which the DRW is defined. The distributed modes are so called because they are distributed over the physical ones, either symmetrically~($+$) or anti-symmetrically~($-$). The mathematical transformation between the two types of modes is equivalent to a 50/50 beamsplitter interaction. 

The logical qumode is defined as the $+$~distributed mode (a.k.a.~the +~macronode subspace), with quadrature operators~$(q_{i+}, p_{i+})$. This is the natural choice because it allows for simple encoding of input states via 50/50 beamsplitter interaction~\cite{ulbttowqc, ulscvcsmittd}. As we shall see, macronode computation on such qumodes bears strong resemblance to CV teleportation, as previously pointed out in Ref.~\cite{ulscvcsmittd}.

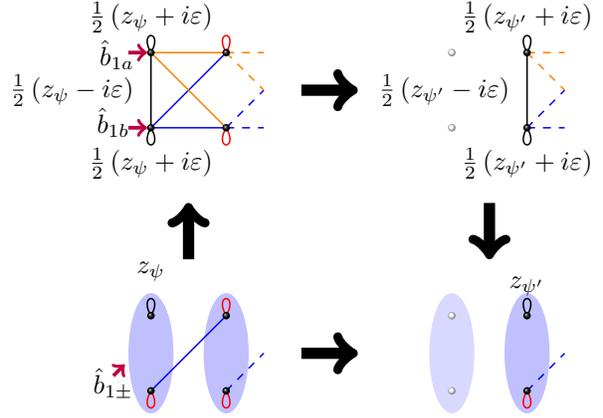
\begin{figure}
\hspace{-4cm}
%
\begin{tikzpicture}[scale=1]
\path (0,0) coordinate (origin);
\path (1,1) coordinate (P0);
\path (1,0) coordinate (P1);
\path (0,1) coordinate (E1);
\path (-0.5, 0.5) coordinate (E2);
%
\path (P0) edge[ out=120, in=60, looseness=1, loop, distance=0.35cm, color =red, line width=0.2mm]
            node[above=3pt] {
} (P0);
\path (P1) edge[ out=-120, in=-60, looseness=1, loop, distance=0.35cm, color =red, line width=0.2mm]
            node[below=3pt] {
} (P1);
\path (origin) edge[ out=-120, in=-60, looseness=1, loop, distance=0.35cm, color = black, line width=0.2mm]
            node[below=3pt] {
} (origin);
\path (E1) edge[ out=120, in=60, looseness=1, loop, distance=0.35cm, color = black, line width=0.2mm]
            node[below=3pt] {} (E1);
 \begin{scope}[color=black,line width=0.2mm]
\draw (origin) -- (E1);
\end{scope}
 \begin{scope}[color=blue,line width=0.2mm]
\draw (origin) -- (P0);
\end{scope}
 \begin{scope}[color=blue,line width=0.2mm]
\draw (origin) -- (P1);
\end{scope}
 \begin{scope}[color=orange,line width=0.2mm]
\draw (E1)--(P0) (E1)--(P1);
\end{scope}
 \begin{scope}[color=blue,line width=0.2mm, dashed]
\draw (1, 0)-- (1.5, 0.5)  (1, 0)-- (1.5, 0);
\end{scope}
 \begin{scope}[color=orange,line width=0.2mm, dashed]
\draw (1, 1)-- (1.5, 0.5) (1, 1)--(1.5, 1);
\end{scope}
 \shade[ball color=black] (P0) circle(0.05);
 \shade[ball color=black] (P1) circle(0.05);
 \shade[ball color=black] (0,0) circle(0.05) node [below=4pt] {$\frac{1}{2}\left(z_{\psi} + i \varepsilon \right)$};
\shade[ball color=black] (0,1) circle(0.05)  node [above=4pt] {$\frac{1}{2}\left(z_{\psi} + i \varepsilon \right)$};
\path (0,0.5) node[left=3pt] {
$\color{black}\frac{1}{2}\left( z_{\psi} - i\varepsilon\right) \color{blue}$
} (-5,1);
\draw[line width=0.5mm, ->, color=purple]  node[label={[xshift=-0.5cm, yshift=-0.4cm]$\color{black}\hat{b}_{1b}\color{blue}$}] {} (-0.3,1) -- (-0.05,1);
\draw[line width=0.5mm, ->, color=purple]  node[label={[xshift=-0.45cm, yshift=0.5cm]$\color{black}\hat{b}_{1a}\color{blue}$}] {} (-0.3,0) -- (-0.05,0);
\draw[line width=1.5mm, ->, color=black] (2,0.5) -- (2.75,0.5);
\draw[line width=1.5mm, ->, color=black] (0.5,-1.75) -- (0.5,-1);
\draw[line width=1.5mm, ->, color=black] (2,-3) -- (2.75,-3);
\draw[line width=1.5mm, ->, color=black] (4.5,-1) -- (4.5,-1.75);
\fill [color=blue, opacity=0.25]  (0 ,-3) ellipse (0.30 and 0.8);
\fill [color=blue, opacity=0.25]  (1,-3) ellipse (0.30 and 0.8);
\path (0, -3.5) coordinate (origin);
\path (1,-2.5) coordinate (P0);
\path (1, -3.5) coordinate (P1);
\path (0, -2.5) coordinate (E1);
\path (0.5, -3) coordinate (E2);
\path (P0) edge[ out=120, in=60, looseness=1, loop, distance=0.35cm, color = red, line width=0.2mm]
            node[above=3pt] {
} (P0);
\path (P1) edge[ out=-120, in=-60, looseness=1, loop, distance=0.35cm, color =red, line width=0.2mm]
            node[below=3pt] {
} (P1);
\path (origin) edge[ out=-120, in=-60, looseness=1, loop, distance=0.35cm, color = red, line width=0.2mm]
            node[below=3pt] {
} (origin);
\path (E1) edge[ out=120, in=60, looseness=1, loop, distance=0.35cm, color =black, line width=0.2mm]
            node[above=3pt] {$\color{black}z_{\psi}\color{blue}$} (E1);
 \begin{scope}[color=blue,line width=0.2mm]
\draw (origin) -- (P0);
\end{scope}
 \begin{scope}[color=blue,line width=0.2mm, dashed]
\draw (1,-3.5) -- (1.5, -3);
\end{scope}
 \shade[ball color=black] (origin) circle(0.05);
 \shade[ball color=black] (P0) circle(0.05);
 \shade[ball color=black] (P1) circle(0.05);
\shade[ball color=black] (E1) circle(0.05);
\draw[line width=0.5mm, ->, color=purple]  node[label={[xshift=-0.5cm, yshift=-3.9cm]$\color{black}\hat{b}_{1\pm}\color{blue}$}] {} (-0.5,-3.3) -- (-0.35,-3.15);
\path (4,0) coordinate (origin);
\path (4,1) coordinate (E1);
\path (5,0) coordinate (P1);
\path (5,1) coordinate (P0);
%
%
\path (P1) edge[ out=-120, in=-60, looseness=1, loop, distance=0.35cm, color = black, line width=0.2mm]
            node[below=3pt] {
} (P1);
\path (P0) edge[ out=120, in=60, looseness=1, loop, distance=0.35cm, color = black, line width=0.2mm]
            node[below=3pt] {} (P0);
 \begin{scope}[color=black,line width=0.2mm]
\draw (P1) -- (P0);
\end{scope}
 \begin{scope}[color=blue,line width=0.2mm, dashed]
\draw (5, 0)-- (5.5, 0.5)  (5, 0)-- (5.5, 0);
\end{scope}
 \begin{scope}[color=orange,line width=0.2mm, dashed]
\draw (5, 1)-- (5.5, 0.5) (5, 1)--(5.5, 1);
\end{scope}
 \shade[ball color = white] (4,0) circle(0.05);
 \shade[ball color = white] (4,1) circle(0.05);
 \shade[ball color=black] (5,0) circle(0.05) node [below=4pt] {$\frac{1}{2}\left(z_{\psi^{\prime}} + i \varepsilon \right)$};
\shade[ball color=black] (5,1) circle(0.05)  node [above=4pt] {$\frac{1}{2}\left(z_{\psi^{\prime}} + i \varepsilon \right)$};
\path (5,0.5) node[left=3pt] {
$\color{black}\frac{1}{2}\left( z_{\psi^{\prime}} - i\varepsilon\right) \color{blue}$
} (-1,1);
\path (4,-3.5) coordinate (origin);
\path (4,-2.5) coordinate (E1);
\path (5,-3.5) coordinate (P1);
\path (5,-2.5) coordinate (P0);
\fill [color=blue, opacity=0.25]  (5 ,-3) ellipse (0.30 and 0.8);
\fill [color=blue, opacity=0.15]  (4 ,-3) ellipse (0.30 and 0.8);
\path (P1) edge[ out=-120, in=-60, looseness=1, loop, distance=0.35cm, color =red, line width=0.2mm]
            node[below=3pt] {
} (P1);
\path (P0) edge[ out=120, in=60, looseness=1, loop, distance=0.35cm, color = black, line width=0.2mm]
            node[below=3pt] {} (P0);
 \begin{scope}[color=blue,line width=0.2mm, dashed]
\draw (5, -3.5)-- (5.5, -3);
\end{scope}
 \shade[ball color = white] (4,-3.5) circle(0.05);
 \shade[ball color = white] (4,-2.5) circle(0.05);
 \shade[ball color=black] (5,-3.5) circle(0.05) node [below] {};
\shade[ball color=black] (5,-2.5) circle(0.05)  node [above=4pt] {$z_{\psi^{\prime}}$};
\end{tikzpicture}
\caption{(Color online) Macronode-based computation applies a logical gate to an input encoded in a macronode by measuring the leftmost macronode. Here we assume that the input state is Gaussian (and represent it by the self-loop weight $z_{\psi}$) so that we can describe macronode measurement using the graphical calculus~\cite{gcfgps}. This is for illustrative purposes only, and the same statements can be made for general input states. The top part of the diagram shows a section of the DRW graph corresponding to the $(a,b)$ macronode decomposition---that is, each node represents a  physical qumode. The pair of local homodyne measurements ($\hat{b}_{1a}$, $\hat{b}_{1b}$) apply the state transformation $z_{\psi}\rightarrow z_{\psi^{\prime}}$ (this transformation is described in more detail in Fig.~\ref{fig:dualrailcircuit}). On the bottom is the equivalent description using distributed modes (see  Eq.~\eqref{eq:distlabels}). Notice that while the measurements in this decomposition are non-local (they are effectively Bell measurements), the input state $z_{\psi}$ and output state $z_{\psi^{\prime}}$ are localized on some graph node. Following either set of arrows from the bottom left to the bottom right then shows the graphical evolution of the input state. Also note that on the bottom pair of figures, the bottom right node is merely a spectator. Thus, we can describe this process using distributed modes by a single-qumode input, a two-qumode cluster state (the diagonally joined pair of nodes on the bottom left graph), and Bell measurements (which correspond to local measurements on the physical qumodes). This highlights the similarity between macronode computation and CV teleportation~\cite{ulscvcsmittd}.} \label{fig:dualflow}
\end{figure}

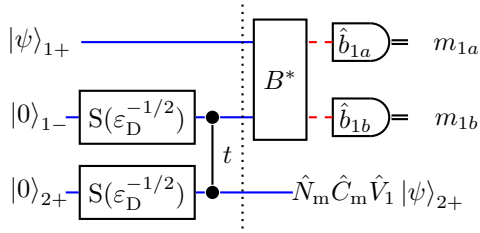
\begin{figure}
\begin{center}
    %
\begin{tikzpicture}[thick]
    %
    \tikzstyle{operator} = [draw,fill=white,minimum size=1.5em] 
    \tikzstyle{phase} = [fill,shape=circle,minimum size=5pt,inner sep=0pt]
    %
 \node at (0,1) (q0) {$\ket{\psi}_{1+}$};
    \node at (0,0) (q1) {$\ket{0}_{1-}$};
    \node at (0,-1) (q2) {$\ket{0}_{2+}$};
    %
    \node[operator] (op20) at (1.3,0) {S($\varepsilon_{\text{D}}^{-1/2}$)} edge [-, blue] (0.35, 0);
  \node[operator] (op21) at (1.3,-1) {S($\varepsilon_{\text{D}}^{-1/2}$)} edge [-, blue] (0.35, -1);
    %
    \node[phase] (phase11) at (2.3,0) {} edge [-, blue] (op20);
    \node[phase] (phase12) at (2.3,-1) {} edge [-, blue] (op21);
    \draw[-] (phase11) -- (phase12);
\node at (2.5, -0.5) {$t$};
    %
\draw (4.4, 0.25)-- (3.9,0.25)--(3.9, -0.25)--(4.4,-0.25);
\draw (4.4,-0.25) arc(-90:90:0.25) [thick];
    \node at (4.2,0) (pm) {$\hat{b}_{1b}$}; 
\node at (4.55,0) {} edge [-,double] (4.9,0);
\node at ( 5.55,-0.05) {$m_{1b}$};
%
\draw (4.4, 1.25)-- (3.9,1.25)--(3.9, 0.75)--(4.4,0.75);
\draw (4.4,0.75) arc(-90:90:0.25) [thick];
    \node at (4.2,1) (pm) {$\hat{b}_{1a}$}; 
\node at (4.55,1) {} edge [-,double] (4.9,1);
\node at (5.55,0.95) {$m_{1a}$};
   \node (end0) at (3.1,1) {} edge [-, blue] (q0);
   \node (end3) at (4.03,1) {} edge [-, red, dashed] (end0);
   \node (end1) at (3.1,0) {} edge [-, blue] (phase11);
   \node (end4) at (4.03,0) {} edge [-, red, dashed] (end1);
    \node (end2) at (3.5,-1) {} edge [-, blue] (phase12);
\node at (4.5,-1) {$\hat{N}_{\text{m}} \hat{C}_{\text{m}} \hat{V}_{1}\ket{\psi}_{2+}$};
 \draw [fill=white ] (2.85,-0.3) rectangle (3.55,1.3);
\node at (3.2, 0.5) {$B^{*}$};
\draw[-,dotted] (2.7, 1.5)--(2.7, -1.5);
      \end{tikzpicture}
\end{center}
\caption{(Color online) Circuit diagram showing a basic element of macronode-based quantum computation. The state immediately after the dotted line is equivalent to a small section of DRW using distributed modes, as shown in Fig.~\ref{fig:dualflow}. In that Figure, $\ket \psi$ is represented by $z_{\psi}$ (assumed Gaussian), and the pair of other modes is equivalent to the leftmost diagonally connected pair of nodes. The box labeled $B^{*}$ does not represent a physical gate. Instead, it is a change from distributed modes (blue solid circuit wires) to physical modes (red dashed circuit wires)---
see Eq.~\eqref{eq:distlabels}. Computation proceeds via local physical homodyne measurements $\hat{b}_{1(a,b)}$. The bottom qumode remains in the distributed-mode basis, which is why the circuit line remains blue. Alternatively, we can interpret this diagram in a different way: If $B^{*}$ is taken to be a physical 50/50 beamsplitter gate, and if the colors and subscript labels are ignored, this diagram shows how to construct a small section of the DRW, inject an input state, and use local macronode measurements to drive MBQC. This reveals that the transformation from physical to distributed modes, Eq.~\eqref{eq:distlabels}, is the same as that of the beamsplitter gate used in DRW construction~\cite{tmcvcsulo}. The squeezing factor acting on the pair of vacuum modes is given by $\varepsilon_{\text{D}}^{-1/2}$, and the effective $\hat C_Z$ interaction strength parameter is $t=\tanh{(2\alpha )}$, where $\alpha$ is the overall squeezing parameter~\cite{tmcvcsulo}. The total operation applied on the logical state is $\hat{N}_{\text{m}} \hat{C}_{\text{m}} \hat{V}_{1}$ (see Eqs.~\eqref{eq:compactnoiseoperator}, \eqref{eq:correctdual}, and \eqref{eq:dualgate}.)}\label{fig:dualrailcircuit}
\end{figure}

To drive computation on the ``$+$'' encoded qumodes, we only need local homodyne measurements on the composite pairs of physical qumodes in the macronode. We refer to such measurements in the following way:
\begin{equation}
\hat{b}_{i(a, b)}=\left( c_{\theta_{a}}\hat{p}_{ia} + s_{\theta_{a}} \hat{q}_{ia}, \;\; c_{\theta_{b}} \hat{p}_{ib} + s_{\theta_{b}} \hat{q}_{ib} \right), \label{eq:macromeasure}
\end{equation} 
where we use the shorthand ${s_\theta \coloneqq \sin \theta}$ and ${c_\theta \coloneqq \cos \theta}$. These measurements are written in terms of local homodyne angles~$\theta_a$ and $\theta_b$, which will be used in what follows. We can describe the effect of these measurements on the input graphically by using both the physical and the distributed modes, as in Fig.~\ref{fig:dualflow}. We see that using distributed modes reduces the description of this process to one that requires only 3 graph nodes. We then represent the logical effect of a macronode measurement as a quantum circuit in Fig.~\ref{fig:dualrailcircuit}.

After a macronode measurement $\hat{b}_{i (a,b)}$ of the input macronode~$i$ with measurement outcomes $m_{i a}$ and $m_{i b}$, the total operation applied to the +~encoded input state, with the result left in macronode~$i+1$ is
\begin{equation}
\ket \psi_{i+} \mapsto \hat{N}_{\text{m}} \hat{C}_{\text{m}} \hat{V}_{i} \ket \psi_{(i+1),+},
\end{equation}
where each of the suboperations---the noise operator~$\hat{N}_{\text{m}}$, the displacement $\hat{C}_{\text{m}}$, and desired unitary gate $\hat{V}_{i}$---are described below. 
 
The noise operator $\hat{N}_{\text{m}}$ involves two applications of $\hat{N}$ (Eq.~\ref{eq:cvnoiseoperator}) separated by a Fourier transform, which means the noise gets added to both the $\hat{q}$ and $\hat{p}$ quadratures. We have that
\begin{equation} \label{eq:compactnoiseoperator}
\hat{N}_{\text{m}}=\hat{N}(\varepsilon_{\text{D}} )  \hat{N}_{\text{p}} \left( \frac{\varepsilon_{\text{D}}}{t^{2}} \right), 
\end{equation}
where $t=\tanh{(2\alpha )}$, and
\begin{align} \label{eq:cvnoiseoperator2}
\hat{N}_{\text{p}} (\varepsilon) &\coloneqq \hat F^\dag \hat{N}(\varepsilon) \hat F \propto\exp \left( \frac{-\varepsilon\hat{p}^{2}}{2 } \right),
\end{align}
which is just $\hat{N}(\varepsilon)$ with its behavior exchanged with respect to position and momentum.
%
%
%
In general, $\hat{N}_{\text{m}}$ adds noise asymmetrically to the quadratures, but in the large-squeezing limit ($t^{2}\to 1$), it is almost symmetric.
%
This is clearly different to the CVW case, where each step introduces noise to one of the quadratures in an alternating fashion (see Eq.~\eqref{eq:cvnoiseoperator}).

The correction operator
\begin{equation}
\hat{C}_{\text{m}} \coloneqq \hat{X}( m_q ) \hat{Z} ( m_p ), \label{eq:correctdual}
\end{equation}
is a phase-space displacement in momentum by~$m_p$ followed by one in position by~$m_q$, where
\begin{align}
m_q &\coloneqq \frac{\sqrt{2}(m_{ib} s_{\theta_{ia}} + m_{ia} s_{\theta_{ib}} )}{t s_{\theta_{i-}}}, \\
m_p &\coloneqq -\frac{t \sqrt{2}(m_{ib} c_{\theta_{ia}} + m_{ia} c_{\theta_{ib}} )}{s_{\theta_{i-}}},
\end{align}
which are written in terms of the sum and difference of the local homodyne angles:
\begin{align}
\theta_{i\pm} &\coloneqq \frac{\theta_{ia} \pm \theta_{ib}}{2}. \label{eq:thetapm}
\end{align}
Both shifts depend on the actual macronode measurement outcomes~$m_{ia}$ and~$m_{ib}$, as well as on the choice of observable $\hat{b}_{(a, b)}$. Notice that this is different from the CVW case, in which the correction is solely a position shift and depends only on the measurement outcome and not on the choice of observable (see Eq.~\eqref{eq:correction}). 

Each macronode measurement implements a gate~$\hat V_i$ dependent on the two parameters $\theta_{i\pm}$:
\begin{equation}
\hat{V}_{i}\coloneqq \hat{S} \left( \frac{1}{t} \right) \hat{R} \left( \theta_{i+} \right)\hat{S} \left( \tan{\theta_{i-}} \right)\hat{R} \left( \theta_{i+} \right) . \label{eq:dualgate}
\end{equation} 
Recall that $\hat{S}$ is a squeezing operator (Eq.~\eqref{eq:squeezegate}), and $\hat{R}$ is a rotation operator (Eq.~\eqref{eq:rotationgate}).
Since each macronode measurement offers twice as many measurement degrees of freedom per site as the CVW protocol, arbitrary single-qumode Gaussian operations can be completed with just two macronode measurements~\cite{ulscvcsmittd} instead of four individual node measurements, using a section of DRW half as long as the CVW required by the CVW protocol. A proof of this is given in the following subsection. 

As a computational unit, a single macronode measurement bears a resemblence to sequential measurement of a pair of CVW qumodes (two CVW protocol measurements). Both procedures offer two measurement degrees of freedom that can be used for gate implementation. Also, they both apply a pair of noise operators and displacements. The noise and displacement operators are separated by a Fourier transform in both cases, which means that noise gets added to both quadratures, and the input states are shifted in both phase-space directions. In order to draw a more quantitative comparison between these protocols, we shall see in the next section how this method compares to the CVW protocol in terms of how much noise is introduced per Gaussian unitary gate. The quantity we compare is the scalar variance, which we derive for the macronode protocol below.

\subsection{Implementing gates} \label{sec:macrogate}
Here show that only two macronode measurements are required in order to apply an arbitrary Gaussian unitary using the macronode protocol.\footnote{For $t=1$ this has the same form as the operation introduced as $M_{\text{tel}}$ in Ref.~\cite{ulbttowqc}. For this reason we have labeled the matrix in Eq.~\eqref{eq:mtel} by the letter $\mathbf{M}$.} This was shown in Ref.~\cite{ulscvcsmittd} for $t=1$, and we generalize the proof to arbitrary $t$ (even though it will be later restricted to ${t=\tanh 2\alpha}$).

Define $\mathbf{R}(\phi )$ and $\mathbf{S}(s)$ to be the symplectic matrix representations of $\hat{R}(\phi )$ (Eq.~\eqref{eq:rotationgate}) and $\hat{S}(s)$ (Eq.~\eqref{eq:squeezegate}) respectively. Let 
\begin{equation}
\mathbf{M}_{i}\coloneqq\mathbf{R}\left( \theta_{i+} \right)\mathbf{S}\left( \tan \theta_{i-} \right)\mathbf{R}\left( \theta_{i+} \right), \label{eq:mtel}
\end{equation}
where $\theta_{i\pm}$ are defined in Eq.~\eqref{eq:thetapm}.
Notice the similarity of the definition of the symplectic matrix~$\mathbf M_i$ to that of the unitary~$\hat V_i$ from Eq.~\eqref{eq:dualgate}. Since $\mathbf{V}_{i}$ is the symplectic representation of the latter, we can write it in terms of~$\mathbf M_i$: 
\begin{equation}
\mathbf{V}_{i}=\mathbf{S}\left( \frac{1}{t}\right) \mathbf{M}_{i}. \label{eq:vsm}
\end{equation}
Furthermore, $\mathbf{V}_{i}\mathbf{S}\left( \frac{1}{t}\right)= \mathbf{S}\left( \frac{1}{t}\right) \mathbf{M}_{i}\mathbf{S}\left( \frac{1}{t}\right)$, and we can incorporate the left- and right-multiplication of $\mathbf M_i$ by a squeezing operation into a change of rotation and squeezing parameters:
\begin{align}
	\mathbf{V}_{i}\mathbf{S}\left( \frac{1}{t}\right) &= \mathbf{S}\left( \frac{1}{t}\right) \mathbf{M}_{i} \mathbf{S}\left( \frac{1}{t}\right) \nonumber \\ 
	&=\mathbf{R}\left( \theta^{\prime}_{i+} \right)\mathbf{S}\left( \tan\theta^{\prime}_{i-} \right)\mathbf{R}\left( \theta^{\prime}_{i+} \right), \label{eq:vseqn}
\end{align}
where 
\begin{equation}
\theta^{\prime}_{i\pm}=\frac{1}{2} \left[ \tan^{-1}{\left(\frac{\sigma_{ia}}{t^{2}}\right)} \pm \tan^{-1}{\left(\frac{\sigma_{ib}}{t^{2}}\right)}\right],
\end{equation}
and ${\sigma_{i(a,b)}=\tan \theta_{i(a,b)}}$.

Arbitrary single-qumode Gaussian unitaries can be decomposed into the form $\hat{R}(\theta )\hat{S}(\eta )\hat{R}(\varphi )$~\cite{ulbttowqc, saair}, whose Heisenberg-picture symplectic representation is just $\mathbf{R}(\theta )\mathbf{S}(\eta )\mathbf{R}(\varphi )$. By using two iterations of $\mathbf{V}_{i}$ and setting $\theta_{1-}=\frac{\pi}{4}$, we have 
\begin{align}\label{eq:simplecompactprotocol} \hspace{-0.8cm}
\mathbf{V}_{2}\mathbf{V}_{1}\bigg|_{\theta_{1-}=\frac{\pi}{4}} \!\!=& \mathbf{V}_{2} \mathbf{S}\left( \frac{1}{t} \right) \mathbf{R}(\theta_{1+})\mathbf{S}(1)\mathbf{R}(\theta_{1+}) \nonumber \\ =& \mathbf{V}_{2} \mathbf{S}\left( \frac{1}{t} \right) \mathbf{R}(2\theta_{1+}),
\end{align}
where we have used that $\mathbf{S}(1)=\mathbf{I}$ and $\mathbf{R}(a)\mathbf{R}(b)=\mathbf{R}(a+b)$. Now, using the definition of $\mathbf{V}_{2}$ in conjuction with Eq.~\eqref{eq:vseqn},
\begin{align} \hspace{-0.8cm}
\mathbf{V}_{2}\mathbf{V}_{1}\bigg|_{\theta_{1-}=\frac{\pi}{4}} \!\!=& \mathbf{R}\left( \theta^{\prime}_{2+} \right)\mathbf{S}\left( \tan \theta^{\prime}_{2-} \right)\mathbf{R}\left( \theta^{\prime}_{2+}\right)\mathbf{R}\left( 2 \theta_{1+}  \right)\nonumber \\=&\mathbf{R}\left( \theta^{\prime}_{2+} \right)\mathbf{S}\left( \tan \theta^{\prime}_{2-} \right)\mathbf{R}\left( \theta^{\prime}_{2+} + 2 \theta_{1+}  \right)\!,
\end{align}
Setting $\theta = \theta^{\prime}_{2+} $, $\eta = \tan \theta^{\prime}_{2-} $ and $\varphi = \theta^{\prime}_{2+} + 2 \theta_{1+}$, we recover the decomposition of an arbitrary single-qumode Gaussian unitary, as required,
\begin{equation} 
\mathbf{V}_{2}\mathbf{V}_{1}\bigg|_{\theta_{1-}=\frac{\pi}{4}} \!\!=\mathbf{R}(\theta )\mathbf{S}(\eta )\mathbf{R}(\varphi ).
\end{equation}
It follows that two macronode measurements are sufficient to implement an arbitrary single-qumode Gaussian unitary.

While the above restriction of setting $\theta_{1-}\!\!=\!\frac{\pi}{4}$ yields a unique and sufficient decomposition for all single-qumode Gaussian unitaries, it is not the optimal choice of homodyne-measurement angles for the purposes of minimizing noise from finite squeezing over all such unitaries. Nevertheless, we use this decomposition for convenience in our proofs for the relative bounds between the noise for each protocol because it provides an upper bound on the true noise for the optimal decomposition of the macronode protocol.

\subsection{Scalar variance for the macronode protocol}
Now we wish to compare the macronode protocol to the CVW protocol in terms of how much noise is introduced per single-qumode Gaussian unitary gate. We will use the Wigner formalism to define the scalar variance for this protocol, which offers a compact description of the noise and its dependence on which measurements are made. Furthermore, it allows the noise analysis and protocol comparison to apply to arbitrary input states (including mixed inputs). Then, in Sec.~\ref{subsec:wignersingle}, we derive relative bounds between the scalar variances for each protocol, establishing a quantitative comparison. 

Consider a small section of the DRW as in Fig.~\ref{fig:dualrailcircuit}. 
The initial state of the DRW in the logical basis has a Wigner function of the form
\begin{equation}
W_{\text{in}} (q_{1\mathbf{+}}, p_{1\mathbf{+}}) W_{\text{CVCS}} (q_{1-}, q_{2+}, p_{1-}, p_{2+}),\label{eq:dualrailinput}
\end{equation}
where $W_{\text{in}}(\mathbf{x})$ is the Wigner function for the single-qumode input state, and
\begin{multline}
\hspace{-0.75cm}W_{\text{CVCS}} (q_{1-}, q_{2+}, p_{1-}, p_{2+})= G_{1/\varepsilon_{\text{D}}} (q_{1-}) G_{1/\varepsilon_{\text{D}}} (q_{2+}) \\ \times G_{\varepsilon_{\text{D}}} (p_{1-}-t q_{2+}) G_{\varepsilon_{\text{D}}} (p_{2+}- t q_{1-})
\end{multline}
is the Wigner function for the two-qumode CVCS.

After measuring $\hat{b}_{1(a,b)}$ in accordance with Fig.~\ref{fig:dualrailcircuit}, applying a displacement $\hat{C}_{\text{m}}^{\dagger}$ to cancel the measurement-dependent displacement (Eq.~\eqref{eq:correctdual}), and then averaging the output-state Wigner function over measurement outcomes, we get an expression for our output $W_{\text{avg}} (\mathbf{x}_{2+} ) $, which is analogous to Eq.~\eqref{eq:cvwirewigneroutput} for the CVW. Define $\mathbf{V}_{i}$ to be the symplectic matrix representation of the Heisenberg action of $\hat{V}_{i}$ (Eq.~\eqref{eq:dualgate}). Then, 
\begin{multline}
\hspace{-0.58cm} W_{\text{avg}} (\mathbf{x}_{2+} ) = \\ \int \! \mathrm{d} \eta_{1} \mathrm{d} \eta_{2}  W_{\text{in}} \left( \mathbf{V}_{1}^{-1}\mathbf{x}_{2+}+\boldsymbol\eta \right)  B_{\mathbf{\Sigma}^{\text{m}}_{1}} (\boldsymbol\eta ),
\end{multline}
where $\boldsymbol\eta=\mathbf{V}_{1}^{-1}\boldsymbol\tau$,  
\begin{equation}
\mathbf{\Sigma}^{\text{m}}_{1}=\mathbf{V}_{1}^{-1} \mathbf{\Sigma}_{*}^{*} \mathbf{V}_{1}^{-T},
\end{equation}
 and
\begin{equation}
\mathbf{\Sigma}^{*}_{*}= \begin{pmatrix} \frac{\varepsilon}{t^{2}} & 0 \\ 0 & \varepsilon \end{pmatrix}.
\end{equation}
The matrix $\frac{1}{2}\mathbf{\Sigma}_{*}^{*}$ is interpreted as the covariance matrix corresponding to a pair of Gaussian convolutions, analogous to Eq.~\eqref{eq:compactnoiseoperator}. 
 
Now we repeat the steps taken in Sec.~\ref{subsec:wignersingle}. By iteration, we can get the general form for the average Wigner function after $n$ measurements ($W_{\text{avg}}^{(n)} (\mathbf{x} )$). The $2n$ convolutions can be simplified down to a single bivariate Gaussian convolution with dummy variables $\kappa_{1}$ and $\kappa_{2}$. We will also apply $\hat{V}^{-1}_{1}\dotsm\hat{V}_{n}^{-1}$, the inverse of the total operation applied in the infinite squeezing limit after $n$ measurements. This results in $W^{(n)}_{\text{undo}} (\mathbf{x}_{n})= W_{\text{avg}}^{(n)} (\tilde{\mathbf{V}}_{n} \mathbf{x}_{n})$, where $\tilde{\mathbf{V}}_{n}=\mathbf{V}_{n} \mathbf{V}_{n-1}\dotsm\mathbf{V}_{1}$. Then,
\begin{equation}
W^{(n)}_{\text{undo}} (\mathbf{x}_{n})=  \int \! \mathrm{d} \bold\kappa_{1} \mathrm{d} \kappa_{2}  W_{\text{in}} \left( \mathbf{x}_{(n+1),+}+\boldsymbol\kappa \right) B_{\mathbf{\Sigma}^{\text{m}}_{n}} (\boldsymbol\kappa ), \label{eq:compactundowigner}
\end{equation}
where $\mathbf{\kappa}=(\kappa_{1}, \kappa_{2})^{T}$, and
\begin{equation}
\mathbf{\Sigma}^{\text{m}}_{n}=\sum^{n}_{i=1} \tilde{\mathbf{V}}_{i}^{-1} \mathbf{\Sigma}^{*}_{*} \tilde{\mathbf{V}}_{i}^{-T}. \label{eq:macronodeSigma}
\end{equation}
Analogous to $\boldsymbol\Sigma_{n}$ in the CVW calculation, $\mathbf{\Sigma}^{\text{m}}_{n}$ reveals how the noise is affected by the number and type of measurement made. Indeed, it is almost exactly the same form as Eq.~\eqref{eq:sigmansingle}, differing by replacing $\mathbf{\Sigma}_{*}$ with $\mathbf{\Sigma}_{*}^{*}$, which has an additional non-zero diagonal entry corresponding to the second Gaussian convolution in the position quadrature (see Eq.~\eqref{eq:compactnoiseoperator}).
By defining
\begin{equation}
SV_{\text{m}}(n) \coloneqq \frac{1}{2}\tr (\mathbf{\Sigma}_{n}^{\text{m}}), \label{eq:macronodeSV}
\end{equation}
we can describe the average effect of noise from finite squeezing from using the macronode protocol after $n$ measurements and compare it to the CVW protocol scalar variance $SV(n)$ that we found earlier in Eq.~\eqref{eq:scalarvariance}.

\subsection{Protocol comparison: noise per Gaussian unitary} \label{subsec:protocolcomparison}
Having derived expressions for the scalar variance of the CVW and macronode protocols after $n$ measurements, we can compare each quantity for the number of measurements required to perform an arbitrary Gaussian unitary. For the CVW protocol this is four node measurements. For the macronode protocol this is two marconode measurements. In each case, we have four degrees of freedom in the measurements to implement a Gaussian unitary, which is described by three degrees of freedom. Thus, we have one degree of freedom $\theta_{\text{free}}$ left to optimize such that the scalar variance is minimized. For each of the protocols, let $\hat{E}$ be the desired Gaussian unitary gate, and define the minimum added noise per Gaussian unitary gate by
\begin{equation}
\mathcal{SV}(n ) \coloneqq \min_{\theta_{\text{free}}}{SV(n)}, \label{eq:minSV}
\end{equation}
in the case of the CVW protocol, and
\begin{equation}
\mathcal{SV}_{\text{m}}(n ) \coloneqq \min_{\theta_{\text{free}}}{SV_{\text{m}}(n)}, \label{eq:minSV}
\end{equation}
in the case of the macronode protocol, where calligraphic font distinguishes this minimized quantity on the left-hand side from the one on the right, which is for a particular gate. It can be evaluated by: 
\begin{enumerate}

\item Solving the constraint equation for three of the four free homodyne angles.

\item Minimizing the corresponding scalar variance function $SV_{(\text{m})}(n)$ over the remaining free homodyne angle. 
\end{enumerate}

For the CVW, we have $n=4$, the constraint equation is $\mathbf{E}= \mathbf{\tilde{U}}_{4}$, where $\mathbf{E}$ is the symplectic representation of Heisenberg action of $\hat{E}$ (Eq.~\eqref{eq:Edef}), and the corresponding homodyne angles are $\theta_{1}$, $\theta_{2}$, $\theta_{3}$, and $\theta_{4}$. For the macronode protocol, we have $n=2$, the constraint equation is $\mathbf{E}= \mathbf{\tilde{V}}_{2}$, and the corresponding homodyne angles are $\theta_{1a}$, $\theta_{1b}$, $\theta_{2a}$, $\theta_{2b}$. 

These quantities represent the minimum noise introduced by finite squeezing per Gaussian unitary gate. Similar work was presented in Ref.~\cite{douowqcfcv} for specific examples of gates implemented with four measurements on CVWs. When applied to the CVW protocol, the procedure outlined above generalizes those results to include arbitrary single-qumode Gaussian unitaries implemented using $n$ CVW node measurements. 

We now present the following bound between the minimum scalar variances for the CVW and macronode protocols. This bound is derived in Appendix \ref{sec:appA}. For any $\hat{E}$, we have that
\begin{equation}
\mathcal{SV}(4)\geq \mathcal{SV}_{\text{m}}(2)+ \frac{3\varepsilon}{t^{2}}, \label{eq:noisebound1}
\end{equation}
where $\mathcal{SV} (4)$ and $\mathcal{SV}_{\text{m}} (2)$ are the minimum scalar variances for implementing the gate $\hat{E}$ using four CVW measurements and two macronode measurements, respectively (Eq.~\eqref{eq:minSV}). Hence, the macronode protocol introduces less noise than the CVW protocol per Gaussian unitary gate. 

In the next section we will discuss an application of the macronode protocol. It is possible, through a restriction of the single-qumode measurements, to retrieve a CVW-like protocol from macronode computation. We call this restriction the \emph{dictionary protocol}. If given CVW-protocol measurements that correspond to a desired gate, the dictionary protocol offers a simple translation to macronode measurements, allowing one to apply the same logic gate using the macronode protocol.

\subsection{Dictionary protocol}\label{subsec:dualraildict}
We showed above how the macronode protocol can be used to implement arbitrary Gaussian unitaries, just like standard MBQC using the CVW. By restricting the allowed local homodyne measurements we can deepen this similarity to the level of how each individual macronode measurement transforms the input state. In other words, we provide a measurement \emph{dictionary} that applies the same gate as the CVW \emph{site for site}. This property provides a direct recipe for adapting CVW measurement protocols for any Gaussian gate or algorithm to a macronode measurement protocol. 

There is only one choice for $\hat{b}_{(a, b)}$ that reduces Eq.~\eqref{eq:dualgate} to a CVW form as in Eq.~\eqref{eq:mbqcunitary}. It is the following restriction: set $\theta_{1 a}=\pi/2$ in Eq.~\eqref{eq:macromeasure}, or equivalently, at each macronode, measure along the basis $\hat{b}_{\text{d}}=(\hat{q}_{a}, \hat{p}_{b}+\sigma \hat{q}_{b})$, where $\sigma =\tan{\theta}$. 
Note that at each macronode, we are restricted to only half the degrees of freedom as in the general macronode protocol. This restriction means that the single-qumode measurements bear a close resemblence to the CVW protocol, where measuring $\hat{q}$ deletes the top part of the DRW (see Fig.~\ref{fig:dualsummary} (b)). The key difference between the CVW and dictionary protocols is the encoding of the input state prior to measurement. In the CVW protocol, the input state is encoded on a single CVW node, whereas for the dictionary protocol, it is encoded in the +~macronode subspace. Therefore, while the physical qumode measurements have the same form, they cannot be said to have the same effect on the input states.

Under this restriction, Eq.~\eqref{eq:dualgate} reduces to
\begin{equation}
\hat{V}_{i}\bigg|_{\theta_{ia}=\frac{\pi}{2}}= \hat{F} \hat{S} (t) \hat{P} (2\sigma_{i-} ) =: \hat{W}_{i} , \label{eq:dictionaryunitary}
\end{equation}
where $t=\tanh{2\alpha }$. This is exactly the form as in Eq.~\eqref{eq:mbqcunitary} for a uniform $g=t$ wire, up to a factor of two in the shearing parameter. However, the noise and correction sub-operations still vary from the CVW case. 

Here, the noise operator is the same as for the general macronode protocol, as defined in Eq.~\eqref{eq:compactnoiseoperator}. Note that this does not mean that they introduce the same amounts of noise per Gaussian unitary. That will depend on the measurement bases and the number of measurements made, both of which will vary with the gate implemented. 

The measurement-dependent displacement $\hat{C}_{\text{d}}$ is given by restricting Eq.~\eqref{eq:correctdual} to the case where $\theta_{a}=\frac{\pi}{2}$,
\begin{equation}
\hat{C}_{\text{d}}= \hat{Z} \left( - t \sqrt{2} m_{a} \right) \hat{X} \left( \frac{\sqrt{2} m_{b} + \sqrt{2} s_{\theta} m_{a}}{t c_{\theta}} \right).
\end{equation}
Contrast the above expression with the CVW case,  where the displacement is simply $\hat{X}\left( \frac{m}{g} \right)$. The displacement for the dictionary protocol involves a displacement in both the position and momentum quadratures, and furthermore, there is a dependence on the measurement basis (since $\hat{C}_{\text{d}}$ depends on $\theta$). For the case where $\theta=0$, the correction operator is similar to the measurement-dependent displacement that occurs in CV quantum teleportation~\cite{tocqv}.

Because of the equivalence between $\hat{U}_{i}$ for the CVW and $\hat{W}_{i}$ for the dictionary protocol, we are free to use previous results that apply to the CVW showing that four such measurements are sufficient for all Gaussian unitaries~\cite{ulbttowqc}. 

Given a particular Gaussian unitary and corresponding measurements on the CVW, the dictionary translation rule is
\begin{equation}
\hat{p} + \sigma_{i-} \hat{q} \mapsto \hat{b}^{\text{d}}_{(a, b)} = \left( \hat{q}_{a} , \hat{p}_{b}+\frac{\sigma_{i-}}{2} \hat{q}_{b} \right),
\end{equation}
where the left hand side corresponds to the CVW and the right hand side corresponds to the dictionary protocol. While this applies the same gate to the input state, the noise and correction operators will not translate so simply. A set of measurement bases that minimize the noise from finite squeezing for a particular Gaussian unitary in the CVW case will not necessarily be the optimal choice for the dictionary protocol (and vice versa). 
 
We can also attempt to ``remodel" away the effective $g=t$ wire to a $g=1$ wire in an analogous way to the CVW protocol. Though limited, we show that such a comparison can still be made. 

Consider the total operation applied to an arbitrary input state $\ket{\psi}$ by taking $n$ dictionary-protocol macronode measurements. If displacements are ignored, it can be arranged into the following form:
 \begin{align}
\prod_{i=1}^{n} \left[ \hat{N}\left(\varepsilon_{\text{D}} \right) \hat{F} \hat{S} \left(t \right) \hat{P} \left(2\sigma_{i} \right) \hat{N} \left(\varepsilon_{\text{D}} \right) \right] \ket{\psi},
\end{align}
where the ordering in the product is decreasing from $n\rightarrow 1$, left to right. Note that the two noise operators $\hat{N}(\varepsilon_{\text{D}} )$ are separated by a Fourier transform and a squeezing operation. Thus, the noise will be added unequally to the quadratures, as in Eq.~\eqref{eq:compactnoiseoperator}.
By commuting squeezing terms to the front and back of each unit (so that they cancel with neighboring terms), we get:
\begin{equation}
\prod_{i=1}^{n} \left[ \hat{S} \! \left( \! \frac{1}{\sqrt{t}}\right) \! \hat{N}\left(\frac{\varepsilon_{\text{D}}}{t} \right) \hat{F} \hat{P} \left( \! \frac{2\sigma_{i}}{t} \! \right) \! \hat{N} \left(\frac{\varepsilon_{\text{D}}}{t} \right)\hat{S} \left( \! \sqrt{t} \right) \right] \! \ket{\psi}.
\end{equation}
The squeezers in the ``bulk" cancel, leaving only squeezing terms from the first and last term:
\begin{multline} \hspace{-0.8cm}
\hat{S} \left(\frac{1}{\sqrt{t}}\right) \hat{N}^{-1} \left( \frac{\varepsilon_{\text{D}}}{t}\right) \left(\prod_{i=1}^{n} \left[ \hat{N}\left(\frac{2 \varepsilon_{\text{D}}}{t} \right) \hat{F} \hat{P} \left(\frac{2 \sigma_{i}}{t} \right) \right] \right) \\ \times  \hat{N} \left(\frac{\varepsilon_{\text{D}}}{t} \right)\hat{S} \left( \sqrt{t} \right) \ket{\psi},
\end{multline}
where
\begin{align}
	\hat{N}^{-1} \left( \frac {\varepsilon_{\text{D}}} {t} \right) \coloneqq \hat{N} \left( -\frac {\varepsilon_{\text{D}}} {t} \right)
\end{align}
(up to renormalization of the final state) is the inverse operation to $\hat{N} (\frac{\varepsilon_{\text{D}}}{t})$, defined only formally in order to reduce the leftmost $\hat{N}\left( \tfrac{2 \varepsilon_{\text{D}}}{t}\right)$ in the following way: $\hat{N}^{-1}\left( \tfrac{\varepsilon_{\text{D}}}{t} \right)\hat{N}\left( \tfrac{2 \varepsilon_{\text{D}}}{t} \right)=\hat{N}\left( \tfrac{\varepsilon_{\text{D}}}{t} \right)$.\footnote{Note that $\hat{N}^{-1} (\varepsilon) \ket \psi$ is not in general a normalizable wavefunction for an arbitrary input state~$\ket \psi$. It is normalizable, however, in the case where $\ket\psi \propto \hat{N} (\delta) \ket \phi$ for some normalizable state~$\ket \phi$ and $\delta > \varepsilon$.}

We can identify the encoding operation $\hat{E}_{\text{d}}\coloneqq \hat{N} \left( \frac{\varepsilon_{\text{D}}}{t} \right)\hat{S} \left( \sqrt{t}\right)$. By including the noise operator it makes the structure of each unit (in the square brackets below) be of the same form as for the CVW protocol:
\begin{equation}
\hat{E}^{-1}_{\text{d}
} \left( \prod_{i=1}^{n} \left[ \hat{N}\left( \frac{2\varepsilon_{\text{D}}}{t} \right) \hat{F} \hat{P} \left( \frac{2\sigma_{i}}{t} \right) \right] \right)  \hat{E}_{\text{d}} \ket{\psi},
\end{equation}
where $\hat{E}^{-1}_{\text{d}}=\hat{S} \left( \frac{1}{\sqrt{t}}\right)\hat{N}^{-1} \left( \frac{\varepsilon}{t} \right)$.
Hence, up to encoding, this is equivalent to a $g=1$ wire with $\frac{2\varepsilon}{t}$ self-loop weights. Recall that the $\tfrac{t}{2}$ weight wire could be remodeled into a weight $g=1$ wire with self-loop weights $\tfrac{2 \varepsilon_{\text{D}}}{t}$ as well (see Eq.~\eqref{eq:noiseremodel}). Then, up to the encoding and decoding relations, the dictionary protocol and the weight-$\tfrac{t}{2}$ CVW introduce similar amounts of noise. 

Even though the dictionary and CVW protocols appear similar based on the above reasoning, it might be possible to bound the scalar variance of one by the other over all Gaussian unitaries. If this were possible, then the protocol that introduced less noise would be the better choice for implementing Gaussian unitary gates. In Appendix \ref{sec:appA} we show by counterexample that it is not possible to derive relative bounds on the scalar variances between these protocols over all Gaussian unitary gates. Thus, they can be said to be roughly equivalent, and neither can be said to be optimal in terms of noise per gate. 

Finally, defining 
\begin{equation}
SV_{\text{d}}(n) \coloneqq SV_{\text{m}}(n)\bigg|_{\theta_{1a}=\frac{\pi}{2}, \dots ,\theta_{n a}=\frac{\pi}{2}}, \label{eq:SVdict}
\end{equation}
we can derive a bound between the minimum noise introduced per Gaussian unitary by the dictionary protocol ($\mathcal{SV}_{\text{d}}(4)$, defined analogously to Eq.~\eqref{eq:minSV}) and the general macronode protocol:
\begin{equation}
\mathcal{SV}_{\text{d}}(4)\geq \mathcal{SV}_{\text{m}}(2)+ \frac{\varepsilon_{\text{D}}(1+2 \sqrt{2}t)}{t^{2}}.
\end{equation}
The proof of this inequality is given in Appendix~\ref{sec:appA}. This inequality shows that in the best case, the noise introduced when applying any Gaussian unitary by the dictionary protocol will always be greater than for the general macronode protocol case. Despite this, the dictionary protocol is still useful because, as its name indicates, it provides a direct, dictionary-like translation from CVW measurement protocols to protocols that can be used on the DRW.

In this section we discussed the key features of the macronode protocol. We demonstrated that it can introduce less noise per gate than the CVW protocol over all Gaussian unitaries. Minimizing the noise from finite squeezing is an important feature of any measurement-based scheme for quantum computing using CVCSs~\cite{ulbttowqc}. Thus, this result shows the importance of considering how measurements are used to implement unitary gates, and it highlights the benefits of a macronode-based approach. Furthermore, we also showed that the macronode protocol saves on resource overhead by only requiring half as much DRW length as the CVW protocol to implement arbitrary single-qumode Gaussian unitaries. We also introduced the dictionary protocol, which acts as a translation rule for running CVW algorithms using macronode-based measurements on the DRW. While it introduced more noise per gate than the macronode protocol, it was found to be roughly equivalent to the CVW protocol while maintaining a deep similarity to the CVW in the structure of the measurements used to implement a given gate. In the next section, we use our analysis of the CVW, macronode, and dictionary protocols to compare their performance as the number of measurement degrees of freedom used to implement each Gaussian unitary is varied.

\section{Application: Number of measurements per gate}\label{sec:numbermeasurements}


In the discussion of measurement protocols above, we have assumed that four measurement degrees of freedom are available per Gaussian unitary gate. This might seem surprising given that an arbitrary single-qumode Gaussian unitary is specified (up to displacements) by three parameters~\cite{ulbttowqc}. Nevertheless, there exists a small set (of measure zero) of single-qumode gates that cannot be achieved by three CVW measurements~\cite{ulbttowqc}. Furthermore, it is claimed without proof in Ref.~\cite{ulbttowqc} that one cannot even get close to these forbidden gates without the noise due to finite squeezing becoming arbitrarily large. In this section we will explore this notion in a systematic way, applying the noise analysis framework from the previous sections to the three- and four-measurement CVW and dictionary protocols. Our analysis shows why protocols using fewer than four measurements are inadequate for implementing arbitrary single-qumode gates, even in some approximate sense.

We must be careful when choosing gates to analyze since large noise can also result from trying to implement a gate with more squeezing than is available in the original CV cluster state, a fact made rigorous in Appendix~\ref{sec:app2}. In order to isolate the effect we wish to show (noise from too few measurements) from high-squeezing noise (which will occur regardless of the number of measurements), we restrict our analysis to phase space rotations~$\hat{R}(\theta )$, which do not contain any squeezing---i.e.,~when decomposed as $\hat{R}(\theta )=\hat{R}(\phi )\hat{S} (\eta) \hat{R}(\varphi)$, the squeezing parameter $\ln\eta = 0$.

%

Generically, we expect that using a larger number of cluster measurements per gate will introduce more noise from finite squeezing~\cite{loqcwgcs, beicvcs, embqcwcvs}. We show that this holds true for some gates. However, we find that for a large class of gates, this intuition breaks: less noise is introduced when implementing with four measurements than with three. This has been pointed out for a few specific gates on a particular cluster state of experimental interest~\cite{douowqcfcv}. Here we show this to be true for a large class of rotation gates implemented by the CVW and macronode protocols.

\blk

\subsection{CVW protocol} \label{subsec:3vs4measure}

On the CVW, the scalar variance after the three-measurement implementation of a rotation gate~$\hat R(\theta)$---called $SV_{\hat{R}(\theta )} (3)$ and defined below---is unique since the constraint $\hat{R}(\theta )=\hat{U}_{3}\hat{U}_{2}\hat{U}_{1}$ uniquely specifies all the measurement degrees of freedom. As there are no degrees of freedom to minimize over,  $SV(3)$ can be used interchangeably with $\mathcal{SV}(3)$ (see Eq.~\eqref{eq:CVWminSV}). Then, using the results of Sec.~\ref{subsec:wignersingle}, we have 
\begin{multline} \label{noise3measurementsequationsingle}
\hspace{-0.8cm} SV_{\hat{R}(\theta )} (3) =\\ \frac{1}{2}\sum_{i=1}^{3} \tr\left[ \left(\tilde{\mathbf{U}}^{3R}_{i}(\theta ) \right)^{-1}  \mathbf{\Sigma}_{*} \left(\tilde{\mathbf{U}}^{3R}_{i}(\theta )\right)^{-T} \right],
\end{multline}
where $\mathbf{U}_{i}^{3R} (\theta )$ is just $\mathbf{U}_{i}$ constrained by $\tilde{\mathbf{U}}_{3} =\mathbf{R}(\theta )$. 

For the four-measurement implementation, the analogous quantity to Eq.~\eqref{noise3measurementsequationsingle} has one free measurement degree of freedom. We denote the minimum scalar variance for rotation gates by calligraphic font, $\mathcal{SV}_{\hat{R}(\theta )}(4)$, where the minimization is over the one free measurement angle~$\theta_{\text{free}}$, as in Eq.~\eqref{eq:minSV}. Then,
\begin{multline}\label{noise4measurementsequationsingle}
\hspace{-0.8cm}\mathcal{SV}_{\hat{R}(\theta )}(4) =\\ \min_{\theta_{\text{free}}}{\frac{1}{2}\sum_{i=1}^{4} \tr\left[ \left(\tilde{\mathbf{U}}_{i}^{4R}(\theta ) \right)^{-1}  \mathbf{\Sigma}_{*} \left(\tilde{\mathbf{U}}_{i}^{4R}(\theta )\right)^{-T} \right]},
\end{multline}
where $\mathbf{U}^{4R}_{i} (\theta )$ is just $\mathbf{U}_{i}$ constrained by $\tilde{\mathbf{U}}_{4}=\mathbf{R}(\theta )$ and such that the scalar variance is minimized. 

\begin{figure}
\includegraphics[width=0.5\textwidth]{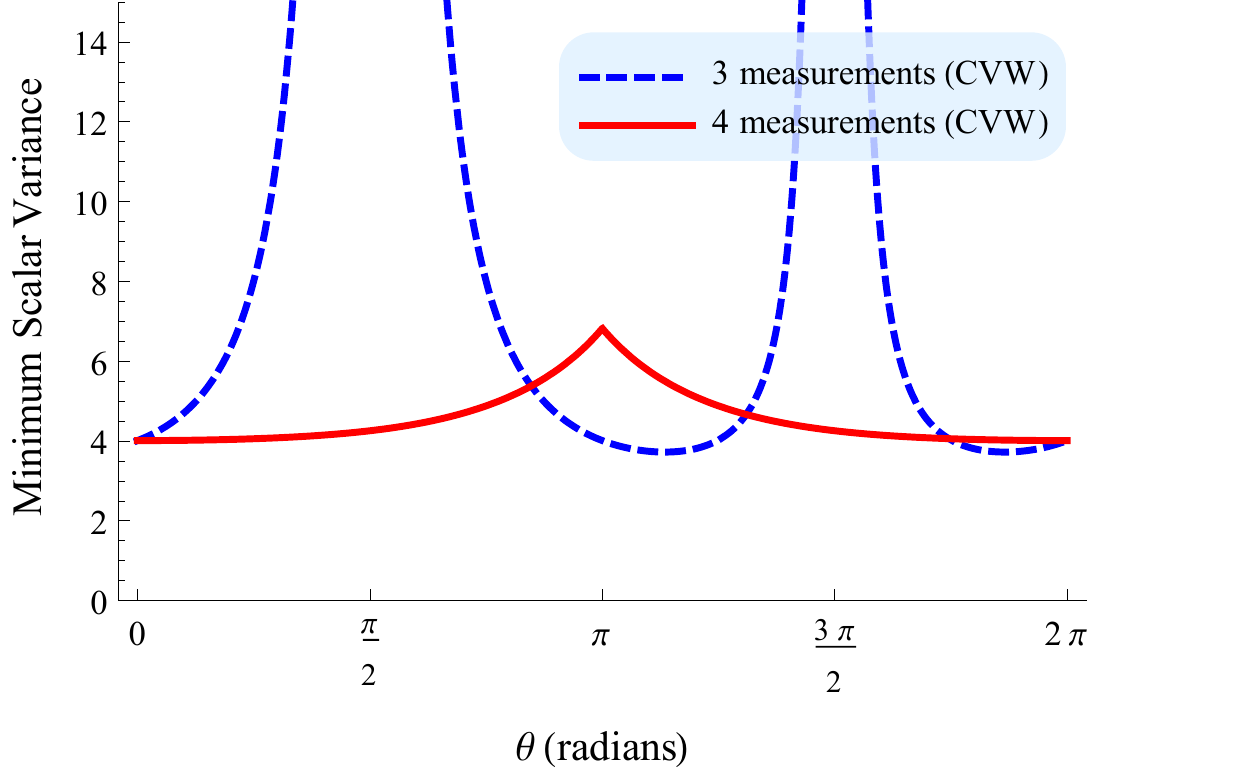}
\caption{(Color online) Minimum scalar variance per rotation gate for three and four measurements on the CVW,  $SV_{\hat{R}(\theta )} (3) $ (dashed blue) and $\mathcal{SV}_{\hat{R}(\theta )}(4)$ (solid red). Technically, only $\mathcal{SV}_{\hat{R}(\theta )}(4)$ has been minimized (represented by the caligraphic font) as $SV_{\hat{R}(\theta )} (3) $ is unique (it has no free measurement degree of freedom). Units on the vertical axis are such that the vacuum variance is $1/2$. Although the three-measurement protocol introduces the least noise at some particular $\theta$, for $\theta$ in the vicinity of $\tfrac{\pi}{2}$ or $\tfrac{3\pi}{2}$, which correspond to gates that cannot be implemented by three CVW measurements~\cite{ulbttowqc}, the noise becomes arbitrarily large. On the other hand, the noise for the four-measurement protocol remains bounded for all $\theta$. In this plot, the squeezing parameter is $\alpha=0.5756$, corresponding to 5~dB of squeezing ($\#~\text{dB} = 10 \log_{10} e^{2\alpha}$), approximately the levels achieved in Ref.~\cite{ulscvcsmittd}.
} \label{singlewirerotationnoise}
\end{figure}

In Fig.~\ref{singlewirerotationnoise} we plot these scalar variances for arbitrary rotations by $\theta$ using three (dashed blue) and four (red) measurements on the CVW. Notice that the noise diverges as a function of angle for the three-measurement case but not for four. This behaviour is generic for all levels of squeezing. In fact, the divergences exactly correspond to those rotation gates that cannot be implemented by the three-measurement CVW protocol~\cite{ulbttowqc}. Note that there are some values of $\theta$ (such as $\theta=\pi$) for which $SV_{\hat{R}(\theta )}(3)<\mathcal{SV}_{\hat{R}(\theta )}(4)$. Thus, there exist instances where applying the three macronode protocol is more efficient than with four measurements---and therefore it could be leveraged to minimize noise further under certain conditions---even though it is clear from Fig.~\ref{singlewirerotationnoise} that this cannot be the general rule.

\subsection{Dictionary protocol}
We can also consider implementing rotations through three and four macronode measurements using the dictionary protocol. Analogous to the CVW protocol case above, we shall denote the scalar variance for a three-measurement implementation of a rotation gate as $SV_{\text{d}, \hat{R}(\theta )}(3)$ by using Eq.~\eqref{eq:SVdict}. As with the CVW protocol, this scalar variance is uniquely determined by the rotation angle $\theta$, and therefore, we can use $SV_{\text{d}}(3)$ and its minimized counterpart $\mathcal{SV}_{\text{d}}(3)$ interchangeably (in the three-measurement case they represent the same quantity). Thus,
\begin{multline}  \label{noise3measurementsequationdual}
\hspace{-0.8cm}SV_{\text{d}, \hat{R}(\theta )}(3)\! = \\ \hspace{-0.3cm} \frac{1}{2}\sum_{i=1}^{3} \tr\!\left[ \left(\tilde{\mathbf{W}}_{i}^{3R}(\theta ) \right)^{-1} \! \mathbf{\Sigma}^{*}_{*} \! \left(\tilde{\mathbf{W}}_{i}^{3R}(\theta )\right)^{-T} \right]\! ,
\end{multline}
where $\tilde{\mathbf{W}}^{3R}_{i} (\theta )$ is just $\tilde{\mathbf{W}_{i}}$ constrained by $\tilde{\mathbf{W}}_{3} =\mathbf{R}(\theta )$, and where $\mathbf{W}_{i}$ is the symplectic matrix representation of the Heisenberg action of $\hat{W}_{i}$. 
Just as in the four-measurement CVW protocol case, the scalar variance for the four-macronode measurement implementation ($SV_{\text{d}, \hat{R}(\theta )}(4) $) has a free measurement degree of freedom. Denote the minimum scalar variance for rotation gates by using calligraphic font, $\mathcal{SV}_{\text{d}, \hat{R}(\theta )}$, where the minimization is over the free measurement degree of freedom ($\theta_{\text{free}}$). Then,
\begin{multline}\label{noise4measurementsequationdual}
\hspace{-0.8cm} \mathcal{SV}_{\text{d},\hat{R}(\theta )}(4) =\\ \min_{\theta_{\text{free}}}{\frac{1}{2} \sum_{i=1}^{4} \tr\left[ \left(\tilde{\mathbf{W}}_{i}^{4R}(\theta ) \right)^{-1}  \mathbf{\Sigma}^{*}_{*} \left(\tilde{\mathbf{W}}_{ i}^{4R}(\theta )\right)^{-T} \right]},
\end{multline}
where the calligraphic font denotes minimization over the free measurement degree of freedom, and $\tilde{\mathbf{W}}^{4 R}_{i} (\theta )$ is just $\tilde{\mathbf{W}}_{i}$ constrained by $\tilde{\mathbf{W}}_{4}=\mathbf{R}(\theta )$ such that the scalar variance is minimized. 

\begin{figure}
\includegraphics[width=0.5\textwidth]{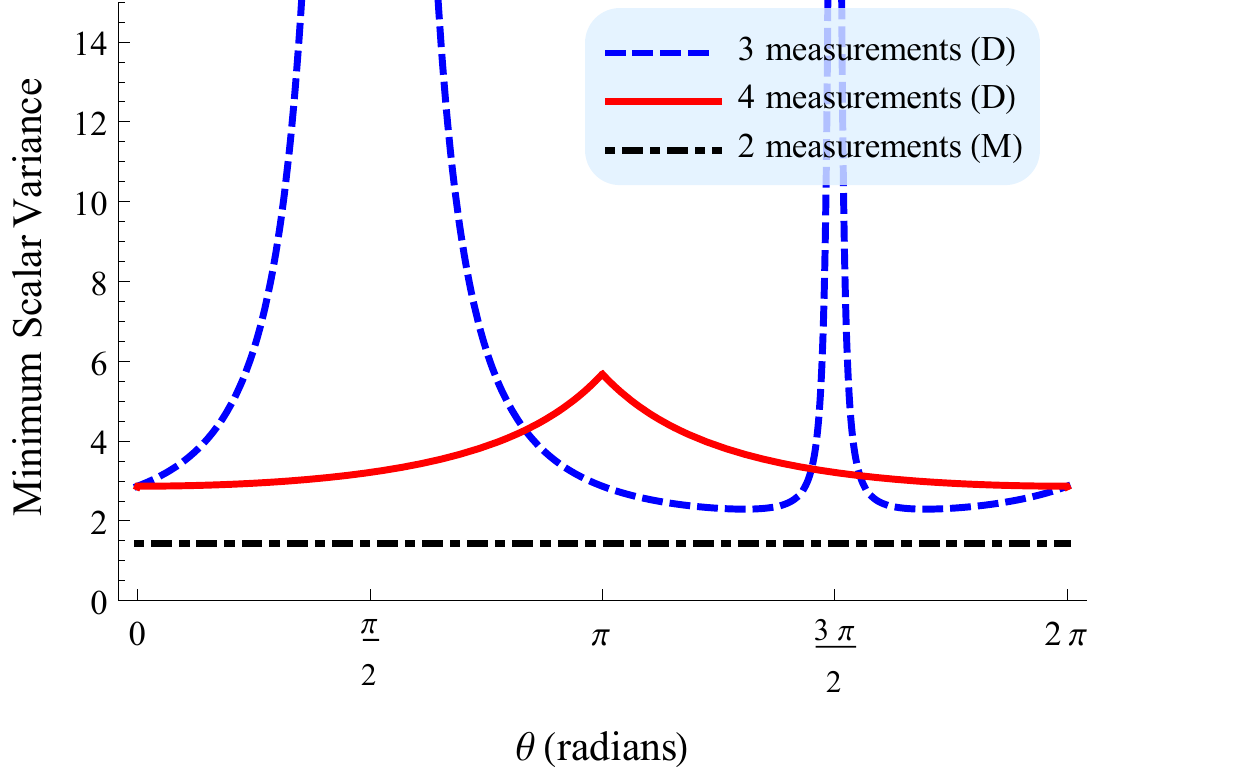}
\caption{(Color online) Minimum scalar variance per rotation gate using the dictionary protocol for three ($SV_{\text{d}, \hat{R}(\theta )}(3)$, dashed blue) and four measurements ($\mathcal{SV}_{\text{d}, \hat{R}(\theta )}(4)$, solid red), as well as the general macronode protocol ($SV^{\text{so}}_{\text{m}, \hat{R} (\theta )}(2))$, dot-dashed black) for two measurements. These are plotted as a function of $\theta$, which specifies the rotation applied to the state. Technically, only $\mathcal{SV}_{\text{d}, \hat{R}(\theta )}(4)$ has been minimized (represented by caligraphic font) as $SV_{\text{d}, \hat{R}(\theta )}(3)$ is unique (it has no free measurement degrees of freedom), and $SV^{\text{so}}_{\text{m}, \hat{R} (\theta )}(2))$ is an upper bound on the minimum for the macronode protocol. Units on the vertical axis are such that the vacuum variance is $1/2$. The variance added by the general macronode protocol is upper-bounded by the black line (corresponding to the suboptimal solution from Eq.~\eqref{eq:simplecompactprotocol}) that does not vary with angle. Although the three-macronode-measurement protocol introduces the least noise at some particular values of~$\theta$, for $\theta$ in the vicinity of $\tfrac{\pi}{2}$ or $\tfrac{3\pi}{2}$, the noise diverges. These angles correspond to gates that cannot be implemented by three CVW measurements~\cite{ulbttowqc} (also expected here because the dictionary protocol is a node-for-node adaptation of the CVW protocol). The noise in the vicinity of these gates becomes arbitrarily large. On the other hand, the noise for the four-measurement protocol remains bounded for all $\theta$. In this plot, the squeezing parameter is $\alpha=0.5756$, corresponding to 5~dB of squeezing ($\#~\text{dB} = 10 \log_{10} e^{2\alpha}$), approximately the levels achieved in Ref.~\cite{ulscvcsmittd}.} \label{dualrailrotationnoise}
\end{figure}

Shown in Fig.~\ref{dualrailrotationnoise} are the scalar variances for rotation gates as a function of angle for three macronode measurements (the dashed blue line) and four macronode measurements (the solid red line) using the dictionary protocol. There is a striking similarity between the three- and four-measurement scalar variance rotation plots in Figs.~\ref{singlewirerotationnoise} and \ref{dualrailrotationnoise}.  Like in the CVW case, the three-measurement dictionary protocol diverges for certain values of $\theta$. In fact, these are the same values of $\theta$ as the ones that have diverging noise in the CVW case. This connection is expected because, as its name implies, the dictionary protocol is a node-for-node mapping of the CVW protocol to macronodes. As such, gates that cannot be applied with three measurements in the CVW case~\cite{ulbttowqc} should similarly fail in the dictionary case, and the noise of both protocols should diverge as one tries to implement gates that are arbitrarily close to them. This is exactly what we see.

Note that there exist other values of~$\theta$ for which ${SV_{\text{d}, \hat{R}(\theta)}(3) < \mathcal{SV}_{\text{d}, \hat{R}(\theta)}(4)}$. Therefore, the three-measurement protocol could be applied in certain cases to minimize the noise per gate further than what is possible with four measurements, even though the existence of divergences for certain angles rules out its use for general single-qumode Gaussian unitaries. 

Also shown in Fig.~\ref{dualrailrotationnoise} (and defined below) is the scalar variance of a two-macronode-measurement implementation of a rotation gate as a function of angle and using the suboptimal shearing parameters from Eq.~\eqref{eq:simplecompactprotocol}  (dot-dashed black line). Denote this as
\begin{multline}
SV^{\text{so}}_{\text{m},  \hat{R}(\theta)} (2) = \\ \frac{1}{2}\sum_{i=1}^{2} \tr\left[ \left(\tilde{\mathbf{V}}^{2 R}_{i}(\theta ) \right)^{-1}  \mathbf{\Sigma}^{*}_{*} \left(\tilde{\mathbf{V}}^{2 R }_{i}(\theta )\right)^{-T} \right]. \label{noise4measurementsequationdualcompact0}
\end{multline}
By the cyclic property of the trace, for all rotations
\begin{equation}
SV^{\text{so}}_{\text{m},  \hat{R}(\theta)} (2) = \tr[\mathbf{\Sigma}^{*}_{*}]. \label{noise4measurementsequationdualcompact}
\end{equation}
By comparing Figs.~\ref{singlewirerotationnoise} and \ref{dualrailrotationnoise}, we observe that the general macronode protocol introduces less than half as much noise for rotation gates as either the CVW or the dictionary protocol using four measurements.  It also outperforms the three-measurement versions of both protocols, albeit by a lesser margin for some angles.

We did not consider the possibility of a three-measurement case for the general macronode protocol here because the measurement degrees of freedom are grouped pairwise per macronode measurement. Therefore, the notion of a three-measurement protocol does not translate clearly.  

In summary, it was claimed without proof in Ref.~\cite{ulbttowqc}---and shown for specific cases in Ref.~\cite{douowqcfcv}---that using four measurements instead of three in a CV cluster-state protocol reduces the overall noise due to finite squeezing because there is an additional measurement degree of freedom that can be adjusted in order to minimize the noise. 
We have shown this to be true for most rotation gates using the CVW and dictionary protocols.
Nevertheless, we have also shown that there are cases where the three-measurement protocol performs better than its four-measurement counterpart, as shown in Figs.~\ref{singlewirerotationnoise} and~\ref{dualrailrotationnoise}. Therefore, three-measurement protocols could be suitable for implementing gates with less noise in certain cases. Finally, we found that implementing rotations with just two measurements using the macronode protocol performs better than either of the other two protocols for both three and four measurements.

\section{Conclusions and Discussion}\label{sec:discussionandconclusion}
In this article we have considered two different approaches to implementing single-qumode Gaussian computation using the dual-rail quantum wire (DRW) resource. We characterized the noise properties of a class of approximate continuous-variable quantum wires (CVWs) in terms of graphical parameters $g$ and $\varepsilon$, which are edge and self-loop weights on the CVW graph respectively~\cite{gcfgps}. We discussed how the value of $g$ affected the logical unitary and noise applied at each measurement and how to modify the measurement protocol on CVWs in order to treat a CVW with uniform weights $\{g, \varepsilon\}$ as if it were one with uniform weights $\{1, \varepsilon g^{-1} \}$ instead. This allows us to parameterize the class of uniform CVWs by a single parameter~$\varepsilon g^{-1}$.

We introduced the macronode protocol for the DRW and proved that it introduces less noise from finite squeezing per single-qumode Gaussian unitary than the CVW protocol. Furthermore, it uses wires of half the length to do so. However, by itself it is by no means a cure for this noise, and it must be combined with other techniques such as active error correction to achieve fault-tolerant quantum computation~\cite{loqcwgcs, beicvcs, embqcwcvs, ftmbqcwcvcs}. Our noise analysis was able to compare a variety of different measurement protocols in terms of the newly defined quantity, the scalar variance $SV(n)=\tfrac{1}{2}\tr[\mathbf{\Sigma}]$. This allowed us to quantify the advantage of using the macronode protocol over the CVW protocol for arbitrary Gaussian unitaries. While we applied this formalism specifically for the choice of parameters given by the temporal-mode linear-optics method~\cite{tmcvcsulo, ulscvcsmittd} and the single-OPO method~\cite{owqcitofc, wqofcicvhcs}, these results are general and can be applied directly to arbitrary weight-$g$ wires over an arbitrary number of linear quadrature measurements.
These results should be extendable to a broader class of continuous-variable cluster states, such as states with a 2D square-lattice graph structure~\cite{tofcaaowqc, wqofcicvhcs} or even higher-dimensional structures such as the hypercubic lattice~\cite{wqofcicvhcs}.

The dictionary protocol provides a theoretical link between the CVW and the DRW, which could potentially be extended to map across other cluster-state features, such as conversion to toric code states~\cite{aswcv, dteeialoqho}. In this Article, we have only considered single-qumode Gaussian unitaries on the DRW, but we anticipate that a similar analysis could be performed for the quad-rail resource state discussed in Ref.~\cite{tmcvcsulo} since homodyne detection on that resource enables multi-qumode Gaussian unitaries. This extension is left to future work. 


\acknowledgments
We thank Natasha Gabay for helpful discussions. This work was supported by the Australian Research Council under grant No.~DE120102204. S.A.~acknowledges financial support from the Prime Minister's Australia Asia Award. S.A.~was supported by the Australian Research Council (ARC) Centre of Excellence for Quantum Computation \& Communication Technology (CQC2T), project number CE110001027. R.U.~acknowledges support from Japan Society for the promotion of Science (JSPS). 


\appendix

\section{Comparative noise bounds between the measurement protocols} \label{sec:appA}
In this appendix we give proofs for the noise bounds claimed in Sec.~\ref{sec:dualrail}. We consider the scalar variances over all single-qumode Gaussian unitaries (up to displacements) for each of the measurement protocols. 

We did not optimize the macronode protocol in deriving these bounds. Instead we used the suboptimal parameters chosen in Eq.~\eqref{eq:simplecompactprotocol} to simplify the calculation. This only provides an upper bound on the minimum macronode-protocol scalar variance.

The first bound we derive compares the CVW protocol and the general macronode protocol.

\subsubsection*{Comparing the CVW and macronode protocols}\label{subsec:singlecompactcompare}
Let the desired Gaussian unitary be denoted by $\hat{E}$. Then define $\mathbf{E}$ to be the sympectic matrix representation of its Heisenberg action.

Using Eq.~\eqref{eq:scalarvariance} and recalling $\mathbf{U}_{i}$ is the symplectic matrix representation of $\hat{U}_{i}$ (Eq.~\eqref{eq:Usymrep}), we expand the scalar variance after four measurements for the CVW protocol:
\begin{multline}
\hspace{-0.75 cm}SV(4)=\frac{1}{2}\tr\left[\tilde{\mathbf{U}}_{4}^{-1}\mathbf{\Sigma}_{*}\tilde{\mathbf{U}}_{4}^{-T}\right] +\frac{1}{2}\tr\left[\tilde{\mathbf{U}}_{3}^{-1}\mathbf{\Sigma}_{*}\tilde{\mathbf{U}}_{3}^{-T}\right] \\ + \frac{1}{2}\tr\left[\tilde{\mathbf{U}}_{2}^{-1} \mathbf{\Sigma}_{*}\tilde{\mathbf{U}}_{2}^{-T}\right]
+\frac{1}{2}\tr\left[\tilde{\mathbf{U}}_{1}^{-1}\mathbf{\Sigma}_{*}\tilde{\mathbf{U}}_{1}^{-T}\right], \label{eq:singlenoiseexpanded}
\end{multline}
where $\tilde{\mathbf{U}}_{i}=\mathbf{U}_{i} \mathbf{U}_{i-1}\dotsm\mathbf{U}_{1}$.
Recall that $\boldsymbol\Sigma^{*}_{*}=\left(\begin{smallmatrix} \frac{\varepsilon}{t^{2}} & 0 \\ 0 & \varepsilon \end{smallmatrix}\right)$ represents the covariance matrix of the bivariate Gaussian convolution that arises in the Wigner-function description of a single macronode measurement.
Recall the definition of $\mathbf{\Sigma}^{*}$ from Eq.~\eqref{eq:sigmaupstarsingle}, and define 
\begin{equation}
T(\mathbf{E}):=\frac{1}{2}\tr{[\mathbf{E}^{-T}\mathbf{\Sigma}_{*}^{*}\mathbf{E}^{-1}]},\label{eq:noises}
\end{equation}
we can expand the first term in Eq.~\eqref{eq:singlenoiseexpanded} by applying the constraint equation $\mathbf{E}=\tilde{\mathbf{U}}_{4}$ and observing that $\boldsymbol\Sigma_{*}^{*}=\boldsymbol\Sigma_{*}+\frac{1}{t^2}\boldsymbol\Sigma^{*}$. Then,
\begin{align}
\hspace{-0.8cm}\frac{1}{2}\tr\left[\tilde{\mathbf{U}}_{4}^{-1}\mathbf{\Sigma}_{*}\tilde{\mathbf{U}}_{4}^{-T}\right]=& T(\mathbf{E})-\frac{1}{2t^{2}}\tr\left[\tilde{\mathbf{U}}_{4}^{-1}\mathbf{\Sigma}^{*}\tilde{\mathbf{U}}_{4}^{-T}\right]\nonumber\\
=& T(\mathbf{E})-\frac{1}{8}\tr\left[\tilde{\mathbf{U}}_{3}^{-1}\mathbf{\Sigma}_{*}\tilde{\mathbf{U}}_{3}^{-T}\right], \label{eq:singlenoiserelation1}
\end{align}
where we have used the following in order to get the second equality:
\begin{multline}\label{singleterm4} \hspace{-1cm}
\tr\left[\tilde{\mathbf{U}}_{4}^{-1}\mathbf{\Sigma}^{*}\mathbf{\tilde{U}}_{4}^{-T}\right]  = \tr\left[\tilde{\mathbf{U}}_{3}^{-1}\mathbf{U}_{4}^{-1}\mathbf{\Sigma}^{*}\mathbf{U}_{4}^{-T}\tilde{\mathbf{U}}_{3}^{-T}\right],
\end{multline}
and 
\begin{align}
\hspace{-0.8cm}\mathbf{U}_{4}^{-1}\mathbf{\Sigma}^{*}\mathbf{U}_{4}^{-T}\!\!\!&=\!\mathbf{P}\!\left(-\sigma_{4} \right)\mathbf{S}\left(\frac{2}{t}\right)\!\mathbf{F}^{-1}\mathbf{\Sigma}^{*}\mathbf{F}\mathbf{S}\left(\frac{2}{t}\right)\! \mathbf{P}\!\left(-\sigma_{4} \right)^{T}\nonumber\\
&=\mathbf{P}\!\left(-\sigma_{4} \right)\mathbf{S}\left(\frac{2}{t}\right)\mathbf{\Sigma}_{*}\mathbf{S}\left(\frac{2}{t}\right)\! \mathbf{P}\!\left(-\sigma_{4} \right)^{T}\nonumber\\
&=\frac{t^{2}}{4} \mathbf{P}\!\left(-\sigma_{4} \right)\mathbf{\Sigma}_{*} \mathbf{P}\!\left(-\sigma_{4} \right)^{T}\nonumber\\
&=\frac{t^{2}}{4}\mathbf{\Sigma}_{*} .
\end{align}
Then, by substituting Eq.~\eqref{eq:singlenoiserelation1} into Eq.~\eqref{eq:singlenoiseexpanded},
\begin{multline}
\hspace{-0.75cm} SV(4)=T(\mathbf{E}) +\frac{3}{8}\tr\left[\tilde{\mathbf{U}}_{3}^{-1}\mathbf{\Sigma}_{*}\tilde{\mathbf{U}}_{3}^{-T}\right] \\  \hspace{-0.75cm}+ \frac{1}{2}\tr\left[\tilde{\mathbf{U}}_{2}^{-1} \mathbf{\Sigma}_{*}\tilde{\mathbf{U}}_{2}^{-T}\right]
+\frac{1}{2}\tr\left[\tilde{\mathbf{U}}_{1}^{-1}\mathbf{\Sigma}_{*}\tilde{\mathbf{U}}_{1}^{-T}\right]. \label{eq:singlenoiseexpanded2}
\end{multline}
Next we minimize the last two terms with respect to the shearing parameters. The minimum occurs when $\sigma_{1}=\sigma_{2}=0$. Then,
\begin{multline}
\hspace{-0.8cm}\text{min}_{\sigma_{1}, \sigma_{2}}\left(\tr\left[\tilde{\mathbf{U}}_{2}^{-1}\mathbf{\Sigma}_{*} \tilde{\mathbf{U}}_{2}^{-T}\right] + \tr\left[\tilde{\mathbf{U}}_{1}^{-1}\mathbf{\Sigma}_{*}\tilde{\mathbf{U}}_{1}^{-T}\right] \right) \\ = \tr\left[\left(\mathbf{S}\left(\frac{2}{t}\right)\mathbf{F}^{-1}\right)^{2}\mathbf{\Sigma}_{*}\left(\mathbf{F}\mathbf{S}\left(\frac{2}{t}\right)\right)^{2}\right]\\+\tr\left[\mathbf{S}\left(\frac{2}{t}\right)\mathbf{F}^{-1}\mathbf{\Sigma}_{*}\mathbf{F}\mathbf{S}\left(\frac{2}{t}\right)\right].
\end{multline}
Then, noting that $\left(\mathbf{S}\left(\frac{2}{t}\right)\mathbf{F}^{-1}\right)^{2}=-\mathbf{I}$ and 
\begin{equation}
\tr\left[\mathbf{S}\left(\frac{2}{t}\right)\mathbf{F}^{-1}\mathbf{\Sigma}_{*}\mathbf{F}\mathbf{S}\left(\frac{2}{t}\right)\right]=\frac{4}{t^{2}}\tr\left[\mathbf{\Sigma}^{*}\right],
\end{equation}
we arrive at
\begin{align} \hspace{-0.75 cm}
\tr\left[\tilde{\mathbf{U}}_{2}^{-1}\mathbf{\Sigma}_{*}\tilde{\mathbf{U}}_{2}^{-T}\right] & + \tr\left[\tilde{\mathbf{U}}_{1}^{-1}\mathbf{\Sigma}_{*}\tilde{\mathbf{U}}_{1}^{-T}\right] \nonumber \\ & \geq\tr\left[\mathbf{\Sigma}_{*}\right]+\frac{4}{t^{2}}\tr\left[\mathbf{\Sigma}^{*}\right] \nonumber \\ & = \tr\left[\mathbf{\Sigma}_{*}^{*}\right]+\frac{3}{t^{2}}\tr\left[\mathbf{\Sigma}^{*}\right]. \label{eq:singlenoiserelation2}
\end{align} 
Now consider the macronode protocol, measuring out 2 macronodes.
Applying the constraint
\begin{equation}
\mathbf{E}=\tilde{\mathbf{V}}_{2}, \label{eq:compactconstraintapp}
\end{equation}
where $\tilde{\mathbf{V}}_{2}=\mathbf{V}_{2}\mathbf{V}_{1}$,
the scalar variance $SV_{\text{m}}(2)$ can be written out explicitly as
\begin{equation} \label{eq:compactnoiseexpand}
SV_{\text{m}}(2)=T(\mathbf{E})+ \frac{1}{2}\tr\left[ \tilde{\mathbf{V}}_{1}^{-1}\mathbf{\Sigma}^{*}_{*} \tilde{\mathbf{V}}_{1}^{-T}\right].
\end{equation}
The minimum scalar variance $\mathcal{SV}_{\text{m}}(2)$ is bounded from above by the scalar variance for the suboptimal solution $SV_{\text{m}}^{\text{so}}(2)$ that uses the choice of homodyne angles in Eq.~\eqref{eq:simplecompactprotocol}. Now,
\begin{align}
\!\!\!\!\!\!\!\!SV^{\text{so}}_{\text{m}}(2) & =\frac{1}{2}\tr{[\mathbf{E}^{-T}\mathbf{\Sigma}_{*}^{*}\mathbf{E}^{-1}]}+\frac{1}{2}\tr{[\mathbf{V}_{1}^{-T}\mathbf{\Sigma}_{*}^{*}\mathbf{V}_{1}^{-1}]}\nonumber \\ &= T(\mathbf{E})+\frac{1}{2}\tr{[\mathbf{\Sigma}_{*}^{*}]} ,\label{eq:noisesubopt}
\end{align}
where we have used the cyclic property of the trace and the fact that $\mathbf{R}^{T}(\theta_{1} )=\mathbf{R}^{-1}(\theta_{1} )$ in the second equality, as well as $\frac{1}{2}\tr{\left[\mathbf{S}\left(\tfrac{1}{t}\right)\mathbf{\Sigma}_{*}^{*}\mathbf{S}\left(\tfrac{1}{t}\right)\right]}=\frac{1}{2}\tr{[\mathbf{\Sigma}_{*}^{*}]}$ and Eq.~\eqref{eq:noises}.
Applying Eqs.~\eqref{eq:noisesubopt} and \eqref{eq:singlenoiserelation2} to Eq.~\eqref{eq:singlenoiseexpanded2}, we have 
\begin{multline}
\hspace{-0.75cm}SV(4)\geq SV_{\text{m}}^{\text{so}}(2)+\frac{3}{2t^{2}}\tr[\mathbf{\Sigma}^{*}] + \frac{3}{8}\tr\left[\tilde{\mathbf{U}}_{3}^{-1}\mathbf{\Sigma}_{*}\tilde{\mathbf{U}}_{3}^{-T}\right].
\end{multline}
As the only constraint that has been placed on the shearing parameters (or equivalently the homodyne angles) is $\tilde{\mathbf{U}}_{4}=\mathbf{E}=\tilde{\mathbf{V}}_{2}$, we are free to replace $SV(4)$ with the scalar variance that has been minimized over the one free measurement angle, $\mathcal{SV}(4)$.
The last term is minimized when $\sigma_{1}=\sigma_{2}=\sigma_{3}=0$. Making a substitution for these parameters yields
\begin{equation}
\mathcal{SV}(4)\geq SV^{\text{so}}_{\text{m}}(2)+\frac{3 \varepsilon_{\text{D}}}{t^{2}},
\end{equation}
from which it follows that
\begin{equation}
\mathcal{SV}(4)\geq \mathcal{SV}_{\text{m}}(2)+\frac{3 \varepsilon_{\text{D}}}{t^{2}} .
\end{equation}
Hence the macronode protocol introduces less noise than the CVW protocol.
\subsubsection*{Comparing the dictionary and general macronode protocols}\label{subsec:dictcompactcompare}
Again, denote the desired Gaussian unitary by $\hat{E}$.
Measuring the $i^{\text{th}}$ macronode on the DRW by the dictionary protocol (in the $\hat{q}_{ia}$ and $\hat{p}_{ib}+\sigma_{i}\hat{q}_{ib}$ bases) applies a Gaussian unitary with symplectic matrix
\begin{equation}
\mathbf{W}_{i}=\begin{pmatrix} -\frac{2\sigma_{i}}{t} & -\frac{1}{t} \\ t & 0 \end{pmatrix}.
\end{equation}
We will consider the scalar variance after four measurements $SV_{d}(4)$ with the requirement that 
\begin{equation}
\tilde{\mathbf{W}}_{4}= \mathbf{E}, \label{eq:dictionaryconstraintapp}
\end{equation}
where $\tilde{\mathbf{W}}_{i}\!\!=\!\!\mathbf{W}_{i}\mathbf{W}_{i-1}\dotsm\mathbf{W}_{1}$, and $\mathbf{E}$ is the symplectic matrix representation of the Heisenberg action of $\hat{E}$.
The scalar variance can be written out explicitly as
\begin{multline}
\hspace{-0.75cm} SV_{\text{d}}(4)=T(\mathbf{E})+\frac{1}{2}\tr\left[\tilde{\mathbf{W}}_{3}^{-1}\mathbf{\Sigma}^{*}_{*}\tilde{\mathbf{W}}_{3}^{-T}\right] \\ \hspace{-1cm}  + \frac{1}{2}\tr\left[\tilde{\mathbf{W}}_{2}^{-1}\mathbf{\Sigma}^{*}_{*}\tilde{\mathbf{W}}_{2}^{-T}\right] 
+ \frac{1}{2}\tr\left[\tilde{\mathbf{W}}_{1}^{-1}\mathbf{\Sigma}^{*}_{*}\tilde{\mathbf{W}}_{1}^{-T}\right],
\end{multline}
Note that
\begin{equation}
\displaystyle \min_{\substack{\sigma_{1 \in \mathbb{R}}}}\tr\left[\tilde{\mathbf{W}}_{1}^{-1}\mathbf{\Sigma}^{*}_{*}\tilde{\mathbf{W}}_{1}^{-T}\right]=\tr{[\mathbf{\Sigma}_{*}^{*}]},
\end{equation}
which is just saying that the minimum noise introduced after one dictionary macronode measurement corresponds to measuring $\hat{q}_{1a}$ and $\hat{p}_{1b}$.

Then, combining the above with Eq.~\eqref{eq:noisesubopt}, and using the suboptimal scalar variance $SV^{\text{so}}_{\text{m}}(2) $ (using suboptimal homodyne angles as in Eq.~\eqref{eq:simplecompactprotocol}) as an upper bound on $SV_{\text{m}}(2)$, we have the following relation:
\begin{multline}
SV_{\text{d}}(4) \geq SV^{\text{so}}_{\text{m}}(2) + \frac{1}{2}\tr\left[\tilde{\mathbf{W}}_{3}^{-1}\mathbf{\Sigma}^{*}_{*}\tilde{\mathbf{W}}_{3}^{-T}\right] \\ +\frac{1}{2}\tr\left[\tilde{\mathbf{W}}_{2}^{-1}\mathbf{\Sigma}^{*}_{*}\tilde{\mathbf{W}}_{2}^{-T}\right].
\end{multline}
Now, what is the minimum value that the last two terms can take? By minimizing those terms over shearing parameters $\sigma_{1}$, $\sigma_{2}$, and $\sigma_{3}$, irrespective of the level of squeezing, 
\begin{multline}
 \frac{1}{2}\!\tr\!\left[\tilde{\mathbf{W}}_{3}^{-1}\mathbf{\Sigma}^{*}_{*}\tilde{\mathbf{W}}_{3}^{-T}\right]  +\frac{1}{2}\!\tr\left[\tilde{\mathbf{W}}_{2}^{-1} \mathbf{\Sigma}^{*}_{*} \tilde{\mathbf{W}}_{2}^{-T}\right]\!  \\ \geq \frac{\varepsilon_{\text{D}}\,(1+2\sqrt{2}t)}{t^{2}}.
\end{multline}
Then, using $SV^{\text{so}}_{\text{m}}(2)\geq \mathcal{SV}_{\text{m}}(2)$, 
\begin{equation}
SV_{\text{d}}(4)\geq \mathcal{SV}_{\text{m}}(2) + \frac{\varepsilon_{\text{D}}\,(1+2\sqrt{2}t)}{t^{2}}
\end{equation}
which is our bound. This bound holds for all dictionary-protocol shearing parameters such that Eqs.~\eqref{eq:dictionaryconstraintapp} and \eqref{eq:compactconstraintapp} are satisfied. In particular it holds for the shearing parameters that correspond the the minimum scalar variance, $\mathcal{SV}_{\text{d}}(4)$. Then,
\begin{equation}
\mathcal{SV}_{\text{d}}(4)\geq \mathcal{SV}_{\text{m}}(2) + \frac{\varepsilon_{\text{D}}\,(1+2\sqrt{2}t)}{t^{2}}.
\end{equation}
Thus, we have proved that the dictionary protocol introduces more noise than the two-measurement macronode protocol over all single-qumode Gaussian unitaries.
\subsubsection*{Comparing the dictionary protocol to the CVW protocol}
 Unfortunately, there is no comparative bound on the noise between the CVW and the dictionary protocol over all Gaussian unitaries. A simple argument for this is to consider a pair of examples. In one case, the dictionary protocol introduces the least noise, and in the other, the CVW protocol does. Consider the identity operation $\mathbf{E}=\mathbf{I}$. The dictionary protocol introduces $\frac{2\varepsilon_{\text{D}} (1+t^{2})}{t^{2}}$ units of noise compared to $\frac{\varepsilon_{\text{D}} (4+t^{2})}{t^{2}}$ for the CVW protocol. In the large-squeezing limit ($t\rightarrow 1$), the dictionary protocol introduces less noise:
\begin{equation}
\text{(dictionary)}\, 4\varepsilon_{\text{D}} \leq 5\varepsilon_{\text{D}} \, \text{(CVW)}.
\end{equation}
For the operation $\mathbf{E}=\mathbf{R}(\pi ) \mathbf{S} (2)$, the dictionary protocol introduces $\frac{55 \varepsilon_{\text{D}}}{8}$ units of noise versus $\frac{\varepsilon_{\text{D}} (5+12 t + 8 t^{2})}{4 t^{2}}$ for the CVW protocol. 
In the large-squeezing limit,
\begin{equation}
\text{(dictionary)}\,\frac{55\varepsilon_{\text{D}}}{8} \geq \frac{25\varepsilon_{\text{D}}}{4}  \, \text{(CVW)},
\end{equation}
and thus, there exist gates for which one protocol is more efficient than the other, and vice versa. This shows that the CVW protocol and the dictionary protocol are roughly equivalent.

%
%

\section{Proof of scalar variance divergences for gates with high squeezing}\label{sec:app2}
Here we prove that, for each of the measurement protocols, any gate that requires high levels of squeezing (relative to that present in the initial CVCS) will introduce a large amount of noise, regardless of how many measurements are used to implement the gate. Let $\hat{E}$ be an arbitrary single-qumode Gaussian unitary, and let $\mathbf{E}$ denote the symplectic matrix representation of its Heisenberg action on the vector of quadrature operators. We can decompose $\mathbf{E}$ as a squeezing matrix~$\mathbf{S}(\eta)$ sandwiched between a pair of rotation matrices~$\mathbf{R}(\theta)$ and~$\mathbf{R}(\varphi )$~\cite{ulbttowqc,saair}:
\begin{equation}
\mathbf{E}=\mathbf{R}(\theta )\mathbf{S}(\eta )\mathbf{R}(\varphi ).
\end{equation}
Consider the scalar variances for the CVW protocol and the macronode protocol (proof for the dictionary protocol is analogous) with respect to an $n$-measurement implementation of the gate~$\hat{E}$. For the CVW protocol, assume that ${n\geq 4}$, which is the minimum number of measurements required to implement any single-qumode Gaussian unitary~\cite{ulbttowqc}. Then,
\begin{align}
SV(n) 
	&=\frac{1}{2}\sum^{n}_{i=1} \tr\left[\tilde{\mathbf{U}}_{i}^{-1} \mathbf{\Sigma}_{*} \tilde{\mathbf{U}}_{i}^{-T}\right].
\end{align}
Note that $\tr\left[\tilde{\mathbf{U}}_{i}^{-1} \mathbf{\Sigma}_{*} \tilde{\mathbf{U}}_{i}^{-T}\right] > 0$ for all~$i$. Therefore, 
\begin{equation}
SV(n) >  \frac{1}{2}\tr\left[ \tilde{\mathbf{U}}_{n-1}^{-1} \mathbf{\Sigma}_{*} \tilde{\mathbf{U}}_{n-1}^{-T}\right] + \frac{1}{2}\tr\left[ \tilde{\mathbf{U}}_{n}^{-1} \mathbf{\Sigma}_{*} \tilde{\mathbf{U}}_{n}^{-T}\right].
\end{equation}
Now, the requirement that the gate $\hat{E}$ is applied by $n$ CVW measurements boils down to demanding that $\tilde{\mathbf{U}}_{n}=\mathbf{E}$. This implies that $\tilde{\mathbf{U}}_{n-1}=\mathbf{U}^{-1}_{n}\mathbf{E}$. Then,  
\begin{align}
SV(n) &> 
	\frac{1}{2}\tr\left[ \mathbf{E}^{-1}\mathbf{U}_{n}\mathbf{\Sigma}_{*} \mathbf{U}^{T}_{n}\mathbf{E}^{-T}\right] \nonumber \\
	& \quad +   \frac{1}{2}\tr\left[ \mathbf{E}^{-1} \mathbf{\Sigma}_{*} \mathbf{E}^{-T}\right].
\end{align}
From Eq.~\eqref{eq:Usymrep} and Eq.~\eqref{eq:sigmaupstarsingle}, we have $\mathbf{U}_{j}\mathbf{\Sigma}_{*}\mathbf{U}^{T}_{j}=g^{-2} \boldsymbol\Sigma^{*}$, which depends only on the uniform self-loop and edge weights, $\varepsilon$ and $g$, respectively. By substitution in the first term for $j=n$, the above inequality becomes 
\begin{align}
	SV(n) &> 
	\frac{1}{2}\tr\left[ \mathbf{E}^{-1} (\mathbf{\Sigma}_{*} + g^{-2}\mathbf{\Sigma}^{*}) \mathbf{E}^{-T}\right]. \label{eq:SVlowerboundapp}
\end{align}

We now turn our attention to the macronode protocol. Assume that ${n\geq 2}$, the minimum number of macronode measurements required to implement any single-qumode Gaussian unitary (see Sec.~\ref{sec:dualrail}). The scalar variance for the macronode protocol is 
\begin{align}
	SV_{\text{m}}(n) 
	&=\frac{1}{2}\sum^{n}_{i=1} \tr\left[\tilde{\mathbf{V}}_{i}^{-1} \mathbf{\Sigma}^{*}_{*} \tilde{\mathbf{V}}_{i}^{-T}\right].
\end{align}
For each~$i$, $\tr\left[\tilde{\mathbf{V}}_{i}^{-1} \mathbf{\Sigma}^{*}_{*} \tilde{\mathbf{V}}_{i}^{-T}\right] > 0$, so
\begin{equation}
SV_{\text{m}}(n) > \frac{1}{2}\tr\left[\tilde{\mathbf{V}}_{n}^{-1} \mathbf{\Sigma}^{*}_{*} \tilde{\mathbf{V}}_{n}^{-T}\right].
\end{equation}
Recall that $\mathbf{\Sigma}^{*}_{*}=\mathbf{\Sigma}_{*}+t^{-2}\mathbf{\Sigma}^{*}$. 
We also require that $\tilde{\mathbf{V}}_{n}=\mathbf{E}$. Thus,
\begin{equation}
SV_{\text{m}}(n) > \frac{1}{2}\tr\left[\mathbf{E}^{-1} (\mathbf{\Sigma}_{*}+t^{-2}\mathbf{\Sigma}^{*}) \mathbf{E}^{-T}\right].
\end{equation}
We have found a lower bound for $SV_{\text{m}}(n)$ that is of the same form as the lower bound for $SV(n)$ in Eq.~\eqref{eq:SVlowerboundapp} (with $t$ in place of $g$)
. We omit the proof for the dictionary protocol since it is identical to the macronode case up to the requirement that ${n\geq 4}$ and setting $\hat{V}_{i}=\hat{W}_{i}$ by the appropriate restriction. 

Now consider this lower bound for general $g$. Denote this quantity as $R(g)$. Using the cyclic property of the trace, we see that
\begin{align}
	R(g) & \coloneqq\frac{1}{2} \tr\left[\mathbf{E}^{-1} (\mathbf{\Sigma}_{*} +g^{-2}\mathbf{\Sigma}^{*}) \mathbf{E}^{-T}\right] \nonumber  \\
	& =\frac{1}{2}\tr\left[ (\mathbf{E}\mathbf{E}^{T})^{-1} (\mathbf{\Sigma}_{*}+g^{-2}\mathbf{\Sigma}^{*})\right]. \label{eq:lowerboundeq} 
\end{align}
Using the decomposition $\mathbf{E}={\mathbf{R}(\theta )\mathbf{S}(\eta )\mathbf{R}(\varphi )}$, note that $\mathbf{E}\mathbf{E}^{T}={\mathbf{R}(\theta )\mathbf{S}(\eta^{2} )\mathbf{R}(-\theta )}$. Therefore, $R(g)$ has no dependence on $\varphi$. Evaluating Eq.~\eqref{eq:lowerboundeq} explicitly yields 
\begin{align}
	R(g) &=\eta^{2}\left(s_{\theta}^{2}\varepsilon + \frac{c_{\theta}^{2}\varepsilon}{g^{2}} \right) + \eta^{-2}\left(c_{\theta}^{2}\varepsilon + \frac{s_{\theta}^{2}\varepsilon}{g^{2}}\right) \nonumber \\
	&\geq \varepsilon (\eta^{2}+\eta^{-2}) \min{\{1, g^{-2}\}}.
\end{align}
Thus, for the CVW protocol,
\begin{equation}
	SV(n) > 
	\begin{cases}
		\phantom{g^{-2}}\varepsilon (\eta^{2}+\eta^{-2}) & \text{if $\abs g \leq 1$,} \\
		g^{-2}\varepsilon (\eta^{2}+\eta^{-2}) & \text{if $\abs g > 1$.} \\
	\end{cases}
\end{equation}
Since ${t=\tanh{2\alpha} <1}$, for the macronode and dictionary protocols, 
\begin{equation}
SV_{(\text{m},\text{d})}(n)> \varepsilon (\eta^{2}+\eta^{-2}). \label{eq:lowerboundDRWsq}
\end{equation}
Therefore, for fixed $\varepsilon$ and $g$, in the large or small limit of $\eta$, the scalar variance for each protocol diverges.  For the CVW case with ${g\leq 1}$ and for the macronode and dictionary protocols, high fidelity in a gate containing squeezing is only possible when ${\varepsilon \ll \min{\{\eta^2, \eta^{-2}\}}}$. When ${g>1}$ in the CVW case, however, a more lenient condition emerges: ${g^{-2} \varepsilon \ll \min{\{\eta^2, \eta^{-2}\}}}$. Thus, increasing the edge weight~$g$ in a CVW to be above~1 may allow for gates with higher squeezing to be implemented with the same~$\varepsilon$. This makes sense in terms of the remodeling protocol of Sec.~\ref{sec:remodelwire}, which showed that increasing $g$ is, in a certain sense, like decreasing~$\varepsilon$. This can be understood by recalling that a higher value of~$g$ represents a stronger $\hat{C}_{Z}$ gate, which itself requires higher squeezing to implement~\cite{saair}. Instead of doing this, one could just redirect that extra squeezing into an effort to decrease~$\varepsilon$ even further. 
%
%

%
%
%
%
%
%
%


\begin{thebibliography}{10}

\bibitem{aowqc}
R.~Raussendorf and H.~J. Briegel, ``A one-way quantum computer'',
\newblock Physical Review Letters {\bf 86}, 5188--5191 (2001).

\bibitem{peiaoip}
H.~J. Briegel and R.~Raussendorf, ``Persistent entanglement in arrays of
  interacting particles'',
\newblock Physical Review Letters {\bf 86}, 910--913 (2001).

\bibitem{cvgaocs}
J.~Zhang and S.~L. Braunstein, ``Continuous-variable gaussian analog of cluster
  states'',
\newblock Physical Review A {\bf 73}, 032318 (2006).

\bibitem{uqcwcvcs}
N.~C. Menicucci {\em et~al.}, ``Universal quantum computation with
  continuous-variable cluster states'',
\newblock Physical Review Letters {\bf 97}, 110501 (2006).

\bibitem{Weedbrook:2012fe}
C.~Weedbrook {\em et~al.}, ``{Gaussian quantum information}'',
\newblock Rev. Mod. Phys. {\bf 84}, 621--669 (2012).

\bibitem{qcwcvc}
M.~L. Gu, C.~Weedbrook, N.~C. Menicucci, T.~C. Ralph, and P.~van Loock,
  ``Quantum computing with continuous-variable clusters'',
\newblock Physical Review A {\bf 79}, 062318 (2009).

\bibitem{bgcsblo}
P.~van Loock, C.~Weedbrook, and M.~Gu, ``Building gaussian cluster states by
  linear optics'',
\newblock Physical Review A {\bf 76}, 032321 (2007).

\bibitem{ugocvcs}
N.~C. Menicucci, S.~T. Flammia, H.~Zaidi, and O.~Pfister, ``Ultracompact
  generation of continuous-variable cluster states'',
\newblock Physical Review A {\bf 76}, 010302 (2007).

\bibitem{owqcitofc}
N.~C. Menicucci, S.~T. Flammia, and O.~Pfister, ``One-way quantum computing in
  the optical frequency comb'',
\newblock Physical Review Letters {\bf 101}, 130501 (2008).

\bibitem{tofcaaowqc}
S.~T. Flammia, N.~C. Menicucci, and O.~Pfister, ``The optical frequency comb as
  a one-way quantum computer'',
\newblock Journal of Physics B-Atomic Molecular and Optical Physics {\bf 42},
  114009 (2009).

\bibitem{alcvcsfasqng}
N.~C. Menicucci, X.~A. Ma, and T.~C. Ralph, ``Arbitrarily large
  continuous-variable cluster states from a single quantum nondemolition
  gate'',
\newblock Physical Review Letters {\bf 104}, 250503 (2010).

\bibitem{tmcvcsulo}
N.~C. Menicucci, ``Temporal-mode continuous-variable cluster states using
  linear optics'',
\newblock Physical Review A {\bf 83}, 062314 (2011).

\bibitem{wqofcicvhcs}
P.~Wang, M.~Chen, N.~C. Menicucci, and O.~Pfister, ``Weaving quantum optical
  frequency combs into continuous-variable hypercubic cluster states'',
\newblock Physical Review A {\bf 90}, 032325 (2014).

\bibitem{eromeo60moaqofc}
M.~Chen, N.~C. Menicucci, and O.~Pfister, ``Experimental realization of
  multipartite entanglement of 60 modes of a quantum optical frequency comb'',
\newblock Physical Review Letters {\bf 112}, 120505 (2014).

\bibitem{ulscvcsmittd}
S.~Yokoyama {\em et~al.}, ``Ultra-large-scale continuous-variable cluster
  states multiplexed in the time domain'',
\newblock Nature Photonics {\bf 7}, 982--986 (2013).

\bibitem{loqcwgcs}
M.~Ohliger, K.~Kieling, and J.~Eisert, ``Limitations of quantum computing with
  gaussian cluster states'',
\newblock Physical Review A {\bf 82}, 042336 (2010).

\bibitem{beicvcs}
H.~Cable and D.~E. Browne, ``Bipartite entanglement in continuous variable
  cluster states'',
\newblock New Journal of Physics {\bf 12}, 113046 (2010).

\bibitem{embqcwcvs}
M.~Ohliger and J.~Eisert, ``Efficient measurement-based quantum computing with
  continuous-variable systems'',
\newblock Physical Review A {\bf 85}, 062318 (2012).

\bibitem{ftmbqcwcvcs}
N.~C. Menicucci, ``Fault-tolerant measurement-based quantum computing with
  continuous-variable cluster states'',
\newblock Physical Review Letters {\bf 112}, 120504 (2014).

\bibitem{Gottesman2001}
D.~Gottesman, A.~Kitaev, and J.~Preskill, ``{Encoding a Qubit in an
  Oscillator}'',
\newblock Phys. Rev. A {\bf 64}, 012310 (2001).

\bibitem{pgoqceitofc}
M.~Pysher, Y.~Miwa, R.~Shahrokhshahi, R.~Bloomer, and O.~Pfister, ``Parallel
  generation of quadripartite cluster entanglement in the optical frequency
  comb'',
\newblock Physical Review Letters {\bf 107}, 030505 (2011).

\bibitem{epoqcaghzesfcv}
X.~L. Su {\em et~al.}, ``Experimental preparation of quadripartite cluster and
  greenberger-horne-zeilinger entangled states for continuous variables'',
\newblock Physical Review Letters {\bf 98}, 070502 (2007).

\bibitem{loqcwpq}
P.~Kok {\em et~al.}, ``Linear optical quantum computing with photonic qubits'',
\newblock Reviews of Modern Physics {\bf 79}, 135--174 (2007).

\bibitem{msmiowqipwsatmol}
R.~Ukai,
\newblock {\em Multi-Step Multi-Input One-Way Quantum Information Processing
  with Spatial and Temporal Modes of Light} (Springer Japan, 2014).

\bibitem{eogcc}
P.~van Loock, ``Examples of gaussian cluster computation'',
\newblock Journal of the Optical Society of America B-Optical Physics {\bf 24},
  340 (2007).

\bibitem{ulbttowqc}
R.~Ukai, J.~I. Yoshikawa, N.~Iwata, P.~van Loock, and A.~Furusawa, ``Universal
  linear bogoliubov transformations through one-way quantum computation'',
\newblock Physical Review A {\bf 81}, 032315 (2010).

\bibitem{gcfgps}
N.~C. Menicucci, S.~T. Flammia, and P.~van Loock, ``Graphical calculus for
  gaussian pure states'',
\newblock Physical Review A {\bf 83}, 042335 (2011).

\bibitem{mtqsol}
U.~Leonhardt,
\newblock {\em Measuring the Quantum State of Light} (Cambridge University
  Press, 1997).

\bibitem{qtpsaoi}
P.~Zanardi, D.~A. Lidar, and S.~Lloyd, ``Quantum tensor product structures are
  observable induced'',
\newblock Physical Review Letters {\bf 92}, 060402 (2004).

\bibitem{saair}
S.~L. Braunstein, ``Squeezing as an irreducible resource'',
\newblock Physical Review A {\bf 71}, 055801 (2005).

\bibitem{douowqcfcv}
R.~Ukai {\em et~al.}, ``Demonstration of unconditional one-way quantum
  computations for continuous variables'',
\newblock Physical Review Letters {\bf 106}, 240504 (2011).

\bibitem{tocqv}
S.~L. Braunstein and H.~J. Kimble, ``Teleportation of continuous quantum
  variables'',
\newblock Physical Review Letters {\bf 80}, 869--872 (1998).

\bibitem{aswcv}
J.~Zhang, C.~D. Xie, K.~C. Peng, and P.~van Loock, ``Anyon statistics with
  continuous variables'',
\newblock Physical Review A {\bf 78}, 052121 (2008).

\bibitem{dteeialoqho}
T.~F. Demarie, T.~Linjordet, N.~C. Menicucci, and G.~K. Brennen, ``Detecting
  topological entanglement entropy in a lattice of quantum harmonic
  oscillators'',
\newblock New Journal of Physics {\bf 16}, 085011 (2014).

\end{thebibliography}


\end{document}